\documentclass[aps,prb,twocolumn,showpacs,superscriptaddress,10pt]{revtex4-1}\begin{tiny}\end{tiny}
\usepackage{longtable}
\usepackage{amsfonts}
\usepackage[dvipdfmx]{graphicx}
\usepackage{rotating}
\usepackage{hyperref}
\usepackage{dcolumn}
\newcolumntype{d}{D{.}{.}{3.5}}

\begin{document}

\title{Insights into the structure of liquid water from nuclear quantum effects on density and compressibility of ice polymorphs}

\author{Bet\"{u}l Pamuk}
 \affiliation{Physics and Astronomy Department, SUNY Stony Brook University, NY 11794-3800, USA}
 \affiliation{School of Applied and Engineering Physics, Cornell University, Ithaca, NY 14853, USA}

\author{P. B. Allen}
 \affiliation{Physics and Astronomy Department, SUNY Stony Brook University, NY 11794-3800, USA}

\author{M-V Fern\'andez-Serra}
 \email{maria.fernandez-serra@stonybrook.edu}
 \affiliation{Physics and Astronomy Department, SUNY Stony Brook University, NY 11794-3800, USA}
 
\date{\today}

%\begin{spacing}{2}

\begin{abstract}

Nuclear quantum effects lead to an anomalous shift of the volume of hexagonal ice; 
heavy ice has a larger volume than light ice.
Furthermore, this anomaly in ice increases with temperature and persists in liquid water up to the boiling point.
To gain more insight, we study nuclear quantum effects on the density and compressibility of several ice-like structures and crystalline ice phases.
By calculating the anisotropic contributions to the stain tensor, 
we analyze how the compressibility changes along different directions in hexagonal ice,
and find that hexagonal ice is softer along the $x$-$y$ plane than the $z$-direction. 
Furthermore, by performing \textit{ab initio} density functional theory calculations with a van der Waals functional
and with the quasiharmonic approximation,
we find an anomalous isotope effect in the bulk modulus of hexagonal ice:
heavy ice has a smaller bulk modulus than light ice.
In agreement with the experiments, we also obtain an anomalous isotope effect for clathrate hydrate structure I.
For the rest of the ice polymorphs, the isotope effect is:
i) anomalous for ice IX, Ih, Ic, clathrate, and low density liquid-like (LDL-like) amorphous ice;
ii) normal at $T=0$ K and becomes anomalous with increasing temperature for ice IX, II, high density liquid-like (HDL-like) amorphous ices, and ice XV;
iii) normal for ice VIII up to the melting point.
There is a transition from an anomalous isotope effect to a normal isotope effect for both the volume and bulk modulus,
as the density (compressibility) of the structures increases (decreases). 
This result can explain the anomalous isotope effect in liquid water:
as the compressibility decreases from melting point to the compressibility minimum temperature, 
the difference between the volumes of the heavy and light water rapidly decreases,
but the effect stays anomalous up to the boiling temperature as the hydrogen bond network is never completely broken by fully filling all the interstitial sites.

\end{abstract}

%\pacs{29.85.-c, 21.10.-k, 23.20.Lv}
\maketitle

%       \newpage
%       \tableofcontents
%       \newpage        

\section{Introduction}

Understanding the structure of ice and water is a challenge especially because
it is difficult to reproduce the experimentally measured anomalies with theoretical simulations.
Upon cooling, the thermal expansion coefficient becomes negative at the temperature of 
maximum density \cite{Kell67} while the specific heat \cite{Smirnova2006} and isothermal compressibility \cite{Kell67}
have a minimum; and all present a divergent behavior in the supercooled region of the phase diagram \cite{Angell1973}.
To explain these anomalous responses in the supercooled region, the existence of a phase transition
between high density and low density liquids with a liquid-liquid critical point
has been hypothesized using empirical force field models \cite{Kumar2008}.
There are many studies in the supercooled \cite{Mallamace2013,Kumar2011,Sciortino2011,Abascal2011} and high temperature \cite{Paschek2004,Pi2009} regimes.
Although some of the simulations find this second critical point \cite{Sciortino2011, Paschek2005, Corradini2010, Abascal2011},
it is not settled that the second critical point is a model independent feature
\cite{Sciortino1997,Moore2011,Limmer2012,Mallamace2013,Nilsson2009,Nilsson2013}.
Furthermore, it is still unclear whether these empirical force field models can capture the correct
physics of hydrogen bonds \cite{Pamuk12}.
Therefore, there is a clear need for an insight from quantum calculations, such as density functional theory
\cite{Corsetti2013}.

\begin{figure}[!htb]
        \centering
                \includegraphics[clip=true, trim=10mm 10mm -10mm -10mm, width=0.45\textwidth]{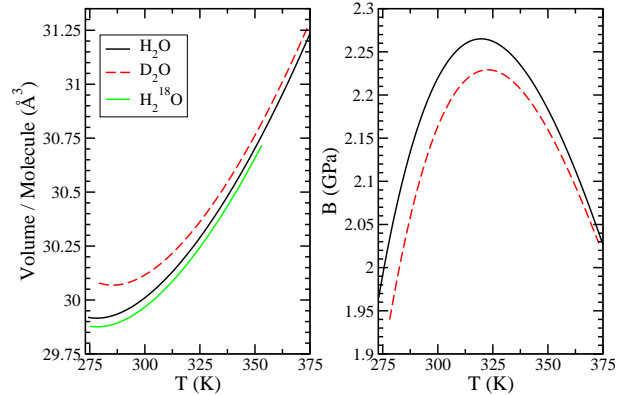}%left bottom right top
        \caption{Volume per molecule (left) and bulk modulus (right) as a function of temperature for light and heavy liquid water
        replotted from Ref. \onlinecite{Kell77}.
		}
        \label{fig:waterV_T}
\end{figure}

Another open question is the contribution of zero point vibrations to the structure of water and ice \cite{Markland2016}.
The temperatures at which anomalies occur are different for normal and heavy ice:
the temperature of specific heat minimum is $310$K ($330$K) \cite{Smirnova2006},
the temperature of maximum density is $277$K ($284$K) \cite{Franks, Kell67},
the temperature of isothermal compressibility minimum is $319.5$K ($322$K) for H$_2$O (D$_2$O) \cite{Kell67}.
Nuclear quantum effects are also observed in the lattice constants of light and heavy hexagonal ice Ih,
as an anomalous isotope effect \cite{Kuhs94}, where the larger volume corresponds to the heavier isotope;
in contrast to the normal isotope effect, where the lighter isotope induces a larger lattice expansion\cite{Allen94}.
Furthermore, this anomalous isotope effect increases up to the melting temperature, and persists to the boiling temperature of liquid water,
as measured in temperature dependence of density and compressibility of light and heavy liquid water \cite{Kell67, Kell77};
as shown in Fig. \ref{fig:waterV_T}, which is replotted using Ref. \onlinecite{Kell77} eq. (34, 38, 41) for volumes and
eq. (35, 39) for compressibility $\kappa$ and bulk modulus $B$, using the relation $B=1/\kappa$.

The primary nuclear quantum effect is revealed in the covalent bond length: $r_{\rm OH}$ is longer than $r_{\rm OD}$.
The secondary nuclear quantum effect, also called the Ubbel\"ohde effect, is revealed in the Hydrogen bond (Hbond) length $r_{\rm OO}$,
the Hbond donor-acceptor (oxygen-oxygen) distance.
Due to the competing quantum effects \cite{Manolopoulos09,Ramesh2014,michaelides11,Zeidler11,Markland2012,Ceriotti2013,Markland2016},
the change in Hbond length depends on the strength of the Hbond:
In a material with a strong Hbond, 
a librational mode (an out-of-plane zero-point motion of the H atom) increases the Hbond length,
while a stretching mode decreases the Hbond length (also called positive Ubbel\"ohde effect).
In a material with a weak Hbond, the opposite effect occurs (negative Ubbel\"ohde effect) .
The Hbond length of ice lies at the crossover between the positive and negative Ubbel\"ohde effect.
To understand the anomalous isotope effect in ice,
we developed a method based on \textit{ab initio} density functional theory (DFT) 
within quasi-harmonic approximation for the volume dependence of the phonon energy.
Using this method, we showed that the anomalous isotope effect is due to the anticorrelations between
the Hbond and OH covalent bond in hexagonal ice \cite{Pamuk12},
reflected in the Gr\"uneisen parameters with opposite signs for librational modes and stretching modes.
With this method, we also showed that 
the nuclear quantum effects are the source of the temperature difference between heavy and normal ice
at the proton order-to-disorder phase transition of hexagonal ices \cite{Pamuk2015}.

In the first section of this paper, we extend our previous studies \cite{Pamuk12, Pamuk2015} to the bulk modulus of hexagonal ices,
to understand the links between the isotope effect and the compressibility.
We analyze the anisotropy in the bulk modulus, 
and the nuclear quantum effects on the isotropic bulk modulus of hexagonal ice,
to show that the anomaly persists in the bulk modulus calculations.

The second section analyzes the anomalous isotope effect of volume and bulk modulus (i.e. compressibility) 
of different phases of ice, to gain insight into the isotope effect in liquid water.
Recent studies show that the quantum effects on the volume are not always anomalous for all phases of ice.
DFT calculations predict a normal isotope effect on the volume of ice VIII \cite{Wentzcovitch10, Galli12, Wentzcovitch2015, Wentzcovitch2017},
which is confirmed by experiments \cite{Wentzcovitch2015}.
In addition, there has not been a study of isotope effects in different phases of ice
that lie in between ice XI and ice VIII. 
Therefore, we analyze different ice-like structures between these two extremes,
with different levels of Hbond, OH covalent bond, and van der Waals (vdW) bond strengths, and interstitial sites.
In addition to the polymorphs: ice XI, ice Ih, ice II, ice IX, ice XV, and ice VIII,
we study clathrate hydrates, where an anomalous isotope effect has been observed experimentally \cite{Kuhs03},
as well as high density liquid-like (HDL-like) and low density liquid-like (LDL-like) amorphous ice structures, 
which resemble liquid water due to the structural disorder in these glassy systems.
With this analysis, we conclude how nuclear quantum effects change with the structure,
and make links between the anomalous isotope effect and the compressibility minimum of liquid water.

\section{Methodology}

\subsection{Ice Structures}

Fig. \ref{fig:ices} shows the ice polymorphs studied in this work. Details of each structure is as follows:

\begin{figure*}[!htb]
        \centering
                \includegraphics[clip=true, trim=0mm 0mm 0mm -10mm, width=1\textwidth]{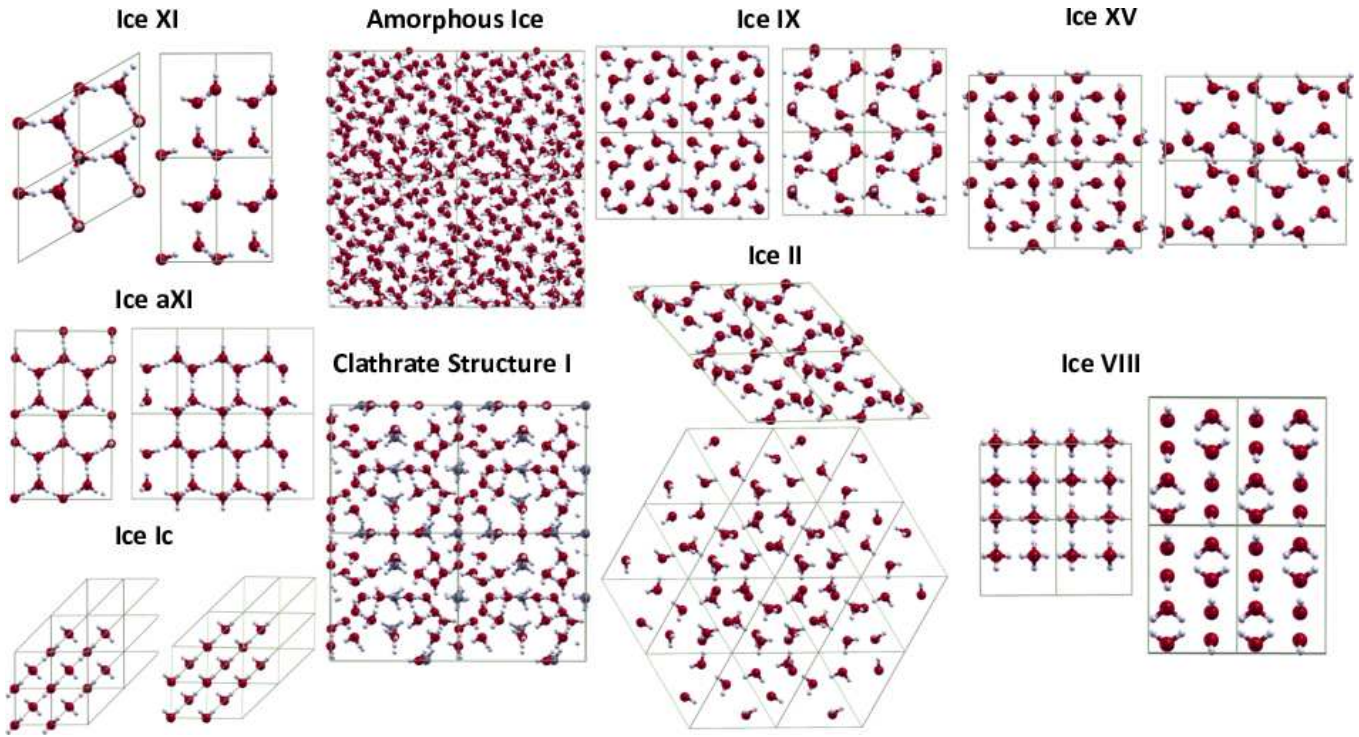}%left bottom right top
        \caption{Ice polymorphs studied in this work. Each structure is shown in a $2\times2\times2$ super cell, and rotated to better show 
        the periodicity of the unit cell.
		}
        \label{fig:ices}
\end{figure*}

\subsubsection{Hexagonal ice}

The most common phase is hexagonal ice Ih, in which oxygen atoms occupy the hexagonal wurtzite structure,
while the hydrogen atoms are disordered. 
The two covalently-bonded protons have six possible orientations, 
but are constrained by Bernal-Fowler ``ice rules'' to have one proton per tetrahedral O-O bond. 
If ice is catalyzed by KOH$^-$ at 72 K, it goes through a phase transition where the protons order \cite{Tajima82}.
Therefore, we consider two possible proton-ordered phases as well as the proton-disordered phase:

(i) Ice XI. This system has 4 molecules in the unit cell. Oxygen atoms are constrained to the hexagonal lattice,
and hydrogen atoms have ferroelectric order with a net dipole moment along the c-axis.

ii) Ice aXI. This system has 8 molecules in the unit cell. The oxygen atoms are ordered as in ice XI,
but the hydrogen atoms have antiferroelectric order with no net dipole moment.

(ii) Ice Ih. Our model for the H-disordered ice Ih structure has 96 molecules in a $3a \times 2\sqrt{3}a \times 2c$ cell.
This system has proton disorder, with no net dipole moment. Both systems have low density liquid like structures.

We have shown before that the order-to-disorder phase transition occurs from the ferroelectric-proton-ordered ice XI,
but for completeness, we also present the antiferroelectric-proton-ordered ice aXI \cite{Pamuk2015}.

\subsubsection{Cubic ice Ic}

Cubic ice Ic is metastable with oxygen atoms are arranged in the cubic diamond structure.
We only study a proton-ordered structure with 2 molecules in the unit cell.

Ice Ic is formed when the temperature of amorphous ice is increased
above $\sim$ 130 K \cite{Blackman1958}.
Alternatively, supercooled water crystallizes to ice Ic upon cooling \cite{Kusalik1994}.
Ice Ic also forms when high pressure phases from ice II to ice IX are recovered in liquid nitrogen and heated \cite{Bertie1963, Bertie1964}.
Therefore, there is no sharp phase transition between ice Ic and ice Ih.
The relationship between the hexagonal ice and the cubic ice is defined by different stacking orders:
ice Ih is AB stacked, while ice Ic is ABC stacked.
There are also studies on the formation of ice with a mixture of ice Ih and ice Ic, with stacking disorder.
We are interested in understanding the isotope effect with changing density.
Therefore the stacking disorder between ice Ic and ice Ih is beyond the scope of this work \cite{Nilsson2017}.

\subsubsection{Clathrate hydrate structure I}

The clathrate hydrate structure I has
two 5$^{12}$ cages and six 5$^{12}$6$^2$ cages for a total of
46 H$_2$O molecules per unit cell \cite{Yndurain10}.
The original structure has a host of 8 CH$_4$ molecules,
but for the purposes of this study, we will focus on the empty clathrate hydrate structure I,
which will be denoted as clathrate sI.

\subsubsection{Amorphous ices}

To approximate amorphous ices, we have taken inherent structures
from different \textit{ab initio} molecular dynamics (AIMD) simulations:

i) One configuration from low density liquid like (LDL-like) structure,
taken from AIMD simulations of 128 water molecules with simulation density $\rho=1$g/cm$^3$
with the PBE functional \cite{Corsetti2013}.

ii) Two configurations from high density liquid like (HDL-like) structure,
taken from AIMD simulations of 64 water molecules with simulation density $\rho=1$g/cm$^3$
with the vdW-DF$^{\rm{PBE}}$ functional \cite{Jue11}.

iii) One configuration from high density liquid like (HDL-like) structure,
taken from AIMD simulations of 128 water molecules with simulation density $\rho=1.2$g/cm$^3$
with the vdW-DF$^{\rm{PBE}}$ functional \cite{Corsetti2013}.

These configurations have both proton and oxygen disorder, 
different from crystalline structures,
resembling the structure of liquid water.

\subsubsection{Ice IX}

Neutron diffraction experiments show that the unit cell of ice IX is tetragonal,
with proton ordering \cite{Londono1993, Kuhs1998},
and with 12 molecules in the unit cell.
Ice IX forms by a phase transition from metastable proton-disordered ice III.
The proton ordering of ice III occurs when high permittivity due to the
proton disorder falls gradually at around 173 K \cite{Whalley1968}.

\subsubsection{Ice II}

The structure of ice II is determined both from X-ray \cite{Kamb1964} and
neutron diffraction measurements \cite{Kamb1971}
to be rhombohedral with $\alpha=113.1^{\circ}$ and with 12 molecules in the unit cell.
The experimental angle is kept constant in all calculations.
This structure is completely proton ordered.
A proton-disordered polymorph of ice II does not exist in the ice phase diagram.
Ice II forms when pressure is applied on hexagonal ice Ih at temperatures between 193 K and 231 K,
or when pressure is released from ice V at 243 K \cite{icebook}.
When ice II is heated, a phase transition to ice III occurs.
However, cooling of ice III does not transform it back to ice II.
Instead, ice III stays metastable until it transforms into ice IX \cite{icebook}.

\subsubsection{Ice XV}

The structure of ice XV is determined to be pseudo-orthorhombic from powder neutron diffraction experiments \cite{Finney2009}.
For simplicity, we used a pseudo-tetragonal cell such that the lattice parameters are $a=b \neq c$, 
with the angles $\alpha=90.1^{\circ}$ and $\beta=\gamma=89.9^{\circ}$ fixed to the experimental values, 
and with 10 molecules in the unit cell.
Ice XV is the antiferroelectric-proton-ordered polymorph of proton-disordered ice VI.
Ice XV is stable at temperatures below $\sim 130$ K and pressures between 0.8 to 1.5 GPa \cite{Finney2009}.

\subsubsection{Ice VIII}

Ice VIII  is determined to be tetragonal from neutron diffraction experiments \cite{Worlton84, Kuhs84},
with 8 molecules in the unit cell, 
and with two interpenetrated networks,
in which molecules of one network occupy the interstitial sites of the other network, 
each with underlying structure of ice Ic.
Within each network, the aforementioned ice rules are satisfied, but there is no Hbond connection between the two networks.
The bonding between these unconnected networks is dominated by the van der Waals interaction.
Each interpenetrating network have ferroelectric order along the $\hat{z}$-axis with opposite signs, resulting in a net
antiferroelectric structure \cite{Kuhs84, Nelmes1993, Besson1994, Nelmes1998}.
Therefore, the inclusion of vdW forces into the calculations become even more important for this case.
Previous calculations of the Hbonded O-O networks and non-bonded O$\cdots$O networks
do not include vdW effects \cite{Wentzcovitch04,Wentzcovitch05,Wentzcovitch2015,Wentzcovitch2017}.
The inelastic neutron scattering spectrum of ice VIII shows a weak coupling between the intramolecular
and intermolecular modes \cite{Eccleston99},
which indeed point to a normal isotope effect in this system.
Ice VIII and its proton-disordered form ice VII are the most dense ice phases, 
located at the high pressure region of the ice phase diagram.

\subsection{Theory}

\subsubsection{Free Energy within Quasiharmonic Approximation}

To include the nuclear quantum effects, the Helmholtz free energy $F(V,T)$ ~\cite{Ziman}
is calculated as a function of volume and temperature in the volume-dependent quasiharmonic approximation (QHA): \cite{Pamuk12,Herrero12}
\begin{eqnarray}
F(V,T) 
  &=& E_0(V) + \nonumber \\
  && \sum_k \left[ \frac{\hbar\omega_k(V) }{2} + 
            k_B T \ln \left(1-e^{-\hbar \omega_k(V) / k_B T}\right) \right].
\label{eq:quasi-harm}
\end{eqnarray}
Here $E_0(V)$ is the energy for classical (or frozen) nuclei, at the relaxed atomic coordinates of each volume.
The volume $V(T)$ and bulk modulus are calculated from the minimum and the curvature of $F(V)$ at each $T$.
The phonon frequencies, $\omega_k$ are calculated with $k$ running over both the branches and the phonon wave vectors within the Brillouin zone.
The volume shift of $\omega_k$ is found for each $T$,
as parameterized by the Gr\"uneisen parameter:
\begin{equation}
\gamma_k=-\frac{\partial({\rm ln} \omega_k)}{\partial({\rm ln} V)} =-\frac{V}{\omega_k} \frac{\partial\omega_k}{\partial V}.
\label{eq:grun}
\end{equation}
In linear approximation this is
\begin{eqnarray}
\omega_k(V) &=& \omega(V_0) \left(\frac{V}{V_0} \right) ^ {-\gamma_k} \\
\label{qh1}
                      & \approx & \omega(V_0) \left( 1- \gamma_k \frac{V-V_0}{V_0} \right).
\label{qh2}
\end{eqnarray}
We have shown previously \cite{Pamuk12, Herrero12}, the QHA as linearized in eq. (\ref{qh2}) is an 
excellent approximation to the full QHA.
Other studies also confirm that the QHA is an appropriate method to analyze the isotope effect on the volume and
volume derivatives of the energy, such as the bulk modulus \cite{Galli12,Wentzcovitch10,Wentzcovitch2015,Wentzcovitch2017,Hirata2016}.

\subsubsection{Bulk Modulus and Pressure}

For the bulk modulus calculations, we change the lattice parameters by $0.15\%$ and 
make a third order polynomial fit to the free energy as a function of volume, $F(V)$ at a given temperature. 
The curvature of this fit gives the bulk modulus,
\begin{equation}
B = V_{\rm min} \left. \frac{\partial^2 F}{\partial V^2} \right|_{V=V_{\rm min}},
\label{eq:B0-E}
\end{equation}
where 
$V_{\rm min}$ is the minimum of the $F(V)$ curve \cite{PhilBulk}.
For the classical (frozen lattice) bulk modulus $B_0$, we use the Kohn-Sham energy $E_0$ instead.

The pressure of the system as a function of volume at a given $T$ is calculated from the free energy
\begin{equation}
P(V) = -\frac{\partial F}{\partial V}.
\label{eq:pres}
\end{equation}

\subsection{Computational Methods}

We use {\sc siesta} code with norm-conserving pseudopotentials,
and numerical atomic basis sets for the valence electrons, for DFT calculations\cite{{SiestaPRBRC},{SiestaJPCM}}.
We use two different density functionals: PBE within generalized gradient approximation (GGA)
to the exchange and correlation, 
and vdW-DF$^{\rm{PBE}}$ with inclusion of long range van der Waals interactions \cite{PBE,DRSLL,SolervdW};
as well as the TTM3F force-field method \cite{ttm3f},
for the bulk modulus calculations of hexagonal ices.
For the rest of the ice-like structures, we present only the results with the vdW-DF$^{\rm{PBE}}$ functional,
since the inclusion of the vdW forces has been shown to be crucial in the analysis of ice polymorphs \cite{Santra2013, Pamuk2015}. 
Comparison of the results with the PBE functional can be found in Ref. \cite{Pamuk2014}.

For hexagonal ices, the same procedure is followed as Ref. \cite{Pamuk12, Pamuk2015}
All initial structural relaxations are done by the t$\zeta$+p atomic orbital basis set for ice Ic, clathrate sI,
and amorphous ices, and 
the d$\zeta$+dp atomic orbital basis set developed in Ref. \cite{Corsetti2013} for ice IX, ice II, ice XV, and ice VIII.
We use a real-space mesh cutoff of 500-600 Ry for the real space integrals,
an electron momentum Monkhorst-Pack grid of $6\times6\times6$ for unit cell calculations,
a force tolerance of 0.005-0.001 eV/\AA~,
and a density matrix tolerance of $10^{-4}$-$10^{-5}$ electrons.
Further details of these parameters for each structure can also be found in Ref. \cite{Pamuk2014}. 
For the volume dependence of the Kohn-Sham energy, the electronic energy of these relaxed configurations
are recalculated using the q$\zeta$+dp atomic orbital basis sets.
The curve for Kohn-Sham energy as a function of volume $E_0(V)$ is obtained and the classical (frozen) volume $V_0$ is
calculated from the minimum of a third order polynomial fit.

For nuclear quantum effects, the vibrational modes are calculated using the frozen phonon approximation.
All force constants calculations are performed with the same basis sets with which the initial relaxations are performed.
We used an atomic displacement $\Delta x$=0.06-0.08 \AA~ for the frozen phonon calculation.
The phonon frequencies, $\omega_k(V_0)$ and Gr\"uneisen parameters $\gamma_k(V_0)$ are obtained by diagonalizing
the dynamical matrix, computed by finite differences from the atomic forces,
at volumes slightly below and above $V_0$.
For crystalline ices, a $3 \times 3 \times 3$ super cell is used.
The Gr\"uneisen parameters are calculated for 3 volumes, and
the phonon modes are calculated by dividing Brillouin zone to a grid of $9\times9\times9$,
with equal weights.

\section{Results}

\subsection{Bulk Modulus of Hexagonal Ices}
\subsubsection{Anisotropy in Classical Bulk Modulus}

We first determine optimal lattice parameters of hexagonal ices.
Lattice parameters are kept constant at each ionic relaxation,
and varied systematically to cover $E_0(a,c)$ surface.
The optimal lattice parameters are selected both considering
the minimum of $E_0(a,c)$ surface, and $E_0(V)$ curve, 
at $P=0$.

\begin{table} [ht] \footnotesize
	\caption{$a$ and $c$ lattice parameters, their ratio ($c/a$), and the corresponding volume per molecule, Volume/H$_2$O$=a^2 c \sqrt{3}/2/N_{{\rm H_2O}}$,
for the force field model (FF) or exchange and correlation functional (XC). All lengths are in \AA~ and volumes in \AA$^3$.}
\centering
%\scalebox{1.0}{
\begin{ruledtabular}
                \begin{tabular} {l l c c c c}
%                \hline
%                \hline
%                \cline{1-6}
                FF/XC               & Ice & $a$  & $c$  & $c/a$   & Volume/H$_2$O \\
                \hline
                TTM3-F              & Ih  & 4.54 & 7.41 & 1.632 & 33.07 \\
                TTM3-F              & aXI & 4.55 & 7.41 & 1.629 & 33.21 \\
                TTM3-F              & XI  & 4.54 & 7.43 & 1.637 & 33.16 \\
                PBE                 & Ih  & 4.42 & 7.21 & 1.631 & 30.49 \\
                PBE                 & aXI & 4.42 & 7.21 & 1.631 & 30.49 \\
                PBE                 & XI  & 4.42 & 7.21 & 1.631 & 30.49 \\
                vdW-DF$^{\rm{PBE}}$ & Ih  & 4.45 & 7.27 & 1.634 & 31.17 \\
                vdW-DF$^{\rm{PBE}}$ & aXI & 4.46 & 7.25 & 1.624 & 31.18 \\
                vdW-DF$^{\rm{PBE}}$ & XI  & 4.45 & 7.25 & 1.629 & 31.08 \\
                Expt. \cite{Kuhs94} 10K $\rm{H_2O}$       & Ih & 4.497 & 7.321 & 1.628 & 32.05 \\% & 0.61 \cite{whalley76}\\
                Expt. \cite{Kuhs94} 10K $\rm{D_2O}$       & Ih & 4.498 & 7.324 & 1.628 & 32.08 \\
                Expt. \cite{Whitworth1996} 5K $\rm{D_2O}$ & Ih & 4.497 & 7.324 & 1.629 & 32.07 \\
                Expt. \cite{Whitworth1996} 5K $\rm{D_2O}$ & XI & 4.501 & 7.292 & 1.620 & 31.98 \\
%                \hline
%                \hline
                \end{tabular}
%               }
\end{ruledtabular}
\label{table:a-c}
\end{table}

Table. \ref{table:a-c} gives calculated lattice parameters of all three hexagonal ice structures,
as well as the experimental results obtained for both heavy and light ice 
by synchrotron radiation \cite{Kuhs94}.
We also present experimental lattice parameters 
from Ref. \onlinecite{Whitworth1996} as they study
neutron diffraction of both ice XI and ice Ih.
They compare the orthorhombic structure of ice XI with the hexagonal structure of ice Ih.
Since we keep hexagonal symmetry in our ice XI calculations, 
we compare 
to experimental lattice parameters $\sqrt{(ab/\sqrt3)}$ of ice XI \cite{Whitworth1996}.
In this case, the experimental lattice parameter $a$ changes
very slightly from ordered to disordered phase,
whereas the lattice parameter $c$ changes significantly.

The TTM3-F force-field model predicts that ice XI has a larger volume than ice Ih,
which is opposite to the experimental results; and ice aXI has the largest volume.
The change is caused by the lattice parameter $a$, when ordering is from ice aXI to ice Ih;
and by $c$, when ordering is from ice XI to ice Ih.
Changes in the $c$ are in opposite directions between the TTM3-F model and the vdW-DF$^{\rm{PBE}}$ functional.
The PBE functional predicts that both proton-ordered ices (ice XI and aXI) have the same lattice parameter as ice Ih.
The vdW-DF$^{\rm{PBE}}$ functional predicts that the antiferroelectric-ordered ice aXI has larger
volume than both ice Ih and ice XI.
The lattice parameter $a$ ($c$) of ice aXI is larger (smaller) than that of ice Ih,
while the difference in the volume of ice aXI and ice XI is only due to the change in the lattice parameter $a$.
Ferroelectric-ordered ice XI has a smaller volume than ice Ih.
This change in volume is due to the change in the lattice parameter $c$,
in good agreement with experiments,
again showing the importance of van-der Waals forces in calculations.
vdW-DF$^{\rm{PBE}}$ 
gives lattice changes similar to experiments between the ferroelectric-ordered ice XI
and proton disordered ice Ih.
This agrees with our previous prediction that the proton order-to-disorder phase transition
is from the ferroelectric-ordered ice XI to ice Ih \cite{Pamuk2015}.

Considering the hexagonal structure of ice, 
with different lattice parameters along different directions as explained above,
the bulk modulus of ice is expected to be different
with respect to the compressions along the $x$-$y$ plane and $z$-axis.
Components of the strain tensor are different for displacements along different directions.
We consider three contributions to strain tensor:

i) $(C_{11}+C_{12})/2$, keeping lattice parameter $c$
constant, while changing $a$ by $0.15\%$.
This shows how the bulk modulus is affected changes along the x-y plane.

ii) $C_{33}$, keeping lattice parameter $a$ constant,
while changing the lattice parameter $c$ by $0.15\%$.
Similarly, this shows how the bulk modulus is affected changes along the z-axis.

iii) $C_{13}$, using the relation between the isotropic bulk modulus 
and other components:
\begin{equation}
B=\frac{C_{33}(C_{11}+C_{12})-2C_{13}^2}{C_{11}+C_{12}+2C_{33}-4C_{13}}
\label{eq:strain_tensor}
\end{equation}

The relations between strain tensor and bulk modulus,
and the derivation of eq. (\ref{eq:strain_tensor}) are explained in detail in the Supporting Information (SI).
The strain tensor components and the bulk modulus results are summarized in Table \ref{table:strain}.
In addition, we check eq. (\ref{eq:strain_tensor}) against the experimental data
of Ref \onlinecite{Gagnon88} in the last two rows of Table \ref{table:strain}.
Using experimental values of B$_0$, ($C_{11}+C_{12}$)/2, and  $C_{33}$, we have calculated $C_{13}$.
The calculated compared reasonably with experiment,
considering the spread both in the experimental and theoretical results.

\begin{table} [htb!] \footnotesize
\caption{The classical bulk modulus and the related components of the strain tensor in units of GPa.
         Experimental results are extrapolated to T=0K from eq. (4) of Ref \onlinecite{Gagnon88}.
         $^e$ is the experimental result from the reference, and $^c$ is the calculated result using eq. (\ref{eq:strain_tensor}).
         }
\centering
%\scalebox{0.85}{
%\begin{center}
\begin{ruledtabular}
                \begin{tabular}{l l c c l c}% c} %p{1.25cm} p{1.25cm} p{1.25cm} p{1.25cm} c c}
%                \hline
%                \hline
%                \cline{1-6}%7}
                FF/XC                  & Ice & B$_0$ & ($C_{11}+C_{12}$)/2 & $C_{33}$ & $C_{13}$ \\  
            \hline
                TTM3-F                 & Ih  & 13.7 & 15.0  & 20.3 & 10.8 \\
                TTM3-F                 & aXI & 14.1 & 14.8  & 20.4 & 12.2 \\
                TTM3-F                 & XI  & 14.2 & 14.4  & 20.8 & 13.3 \\
                PBE                    & Ih  & 14.2 & 17.4  & 27.1 & ~7.9 \\
                PBE                    & aXI & 14.3 & 17.5  & 26.9 & ~8.0 \\
                PBE                    & XI  & 14.3 & 17.4  & 26.3 & ~8.2 \\
                vdW-DF$^{\rm{PBE}}$    & Ih  & 13.5 & 15.7  & 21.9 & ~9.3 \\
                vdW-DF$^{\rm{PBE}}$    & aXI & 14.1 & 16.2  & 20.8 & 10.4 \\
                vdW-DF$^{\rm{PBE}}$    & XI  & 14.3 & 16.1  & 22.1 & 10.6 \\
                Expt \cite{Gagnon88}   & Ih  & ~~~8.48 & ~~10.31  & 14.76 & ~~~~~5.63$^e$  \\
                Expt \cite{Gagnon88}   & Ih  & ~~~8.48 & ~~10.31  & 14.76 & ~~~~~5.09$^c$  \\
%                \hline
%                \hline
                \end{tabular}
\end{ruledtabular}                
%               }
%\end{center}
\label{table:strain}
\end{table}

The experimental results spread over a range between 7-12 GPa,
for the bulk modulus of proton disordered ice Ih.
Few authors comment on the anisotropic components \cite{Wagner06}.
The experimental results \cite{Gagnon88} in Table \ref{table:strain}
lie on the lower end of this spread when isotropic bulk moduli are compared \cite{Gagnon88,Wagner06}.
However, they still give a qualitative intuition about the order of magnitude 
of different components of the strain tensor.
Comparing our calculations of anisotropic bulk modulus along different directions,
shows that bulk modulus along the xy-plane, $(C_{11} + C_{12})/2$,
is smaller than that along the z-axis, ($C_{33}$).
Hexagonal ice is softer along the x-y plane than along the z-axis.

Without a bulk modulus experiment on proton-ordered ice XI
it is difficult to comment on structural differences between ice XI and ice Ih.
The TTM3-F force field model gives no net trend between different components of the strain tensor, 
and an inconsistency between lattice parameter and bulk modulus.
Focusing on the structural differences between the ferroelectric-ordered ice XI
and proton disordered ice Ih, with the semilocal PBE functional, 
all components of the strain tensor are similar.
When non-local vdW forces are included using the vdW-DF$^{\rm{PBE}}$ functional,
both the isotropic and anisotropic moduli of ice XI are clearly larger than those of ice Ih.
This predicts that disordered ice Ih is softer than ordered ice XI.
The vdW-DF$^{\rm{PBE}}$ functional gives more reasonable bulk modulus results
that are consistent with the trend in the volume and lattice parameters.

\subsubsection{Nuclear Quantum Effects on Bulk Modulus}

\begin{figure}[htb]
        \centering
                \includegraphics[clip=true, trim=3mm 13mm -3mm -13mm, width=0.45\textwidth]{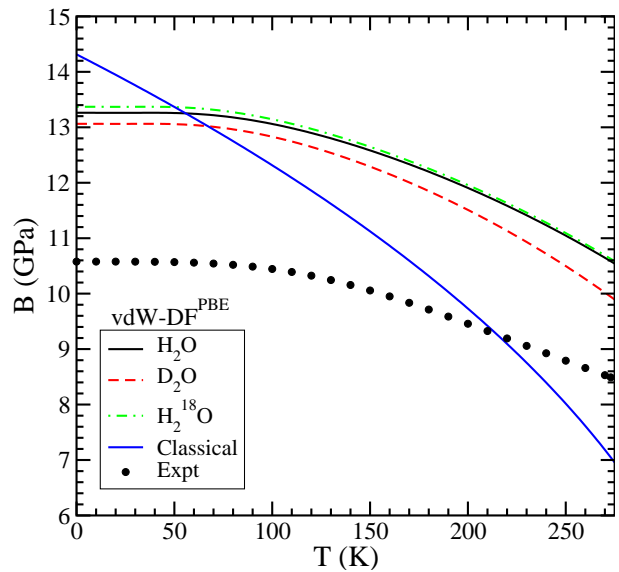}%left bottom right top
        \caption{Bulk modulus as a function of temperature calculated with the QHA using the
        vdW-DF$^{\rm{PBE}}$ functional 
	for proton-ordered ice XI.
        The experimental results are taken from Ref. \onlinecite{Wagner06}.}
        \label{fig:B_T}
\end{figure}

\begin{table*} [htb!] \footnotesize
\caption{The bulk modulus of hexagonal ices including the quantum zero point effects in units of GPa,
			as well as the isotope effect on the bulk modulus IS(A-B)=$\frac{B({\rm A})}{B({\rm B})}-1$ between the isotopes A and B.
			}
%\begin{center}
\centering
\begin{ruledtabular}
                \begin{tabular}{l l c c c c c c}
%                \hline
%                \hline
%                \cline{1-8}
                FF/XC                           & Ice & B$_0$ & H$_2$O & D$_2$O & H$_2\text{}^{18}$O & IS(H-D) & IS($^{16}$O-$^{18}$O)\\
                \hline
                TTM3-F                          & Ih  & 13.74 & 12.81 & 12.85 & 12.86 & $-0.31\%$ & $-0.39\%$ \\
                TTM3-F                          & aXI & 14.13 & 13.03 & 13.08 & 13.09 & $-0.38\%$ & $-0.46\%$ \\
                TTM3-F                          & XI  & 14.24 & 13.04 & 13.09 & 13.10 & $-0.38\%$ & $-0.46\%$ \\
                PBE                             & Ih  & 14.24 & 14.18 & 13.44 & 14.33 & $+5.22\%$ & $-1.06\%$ \\
                PBE                             & aXI & 14.31 & 13.79 & 13.15 & 13.95 & $+4.63\%$ & $-1.17\%$ \\
                PBE                             & XI  & 14.27 & 13.70 & 13.18 & 13.84 & $+3.80\%$ & $-1.02\%$ \\
                vdW-DF$^{\rm{PBE}}$             & Ih  & 13.52 & 13.45 & 13.43 & 13.46 & $+0.15\%$ & $-0.07\%$ \\
                vdW-DF$^{\rm{PBE}}$             & aXI & 14.05 & 13.36 & 13.23 & 13.43 & $+1.00\%$ & $-0.53\%$ \\
                vdW-DF$^{\rm{PBE}}$             & XI  & 14.32 & 13.26 & 13.06 & 13.37 & $+1.51\%$ & $-2.37\%$ \\
                Expt. \cite{hamada10,Gammon83}  & Ih  & & 12.1 \\
                Expt. \cite{Gagnon88}           & Ih  & & ~8.48 \\
                Expt. \cite{Ramirez11,Wagner06} & Ih  & & 10.9 \\
%                \hline
%                \hline
                \end{tabular}
\end{ruledtabular}                
%\end{center}
\label{table:Bulk}
\end{table*}

We also estimate the contributions from quantum zero point effects by calculating the curvature of free energy calculations in eq. (\ref{eq:quasi-harm}).
We computed the phonon density of states and corresponding
Gr\"uneisen parameters, $\gamma_k$, of all structures, as described in eq. (\ref{eq:grun}).
Accurate estimation of Gr\"uneisen parameter is crucial, 
because the average value $\langle \omega_k \gamma_k \rangle$ determines 
whether a system has a normal or anomalous isotope effect \cite{Pamuk12}.
In the normal isotope effect, the bulk modulus of the heavier isotope is larger,
whereas in the anomalous isotope effect, the bulk modulus of the lighter isotope is larger.
Negative $\gamma_k$'s imply a softening of the modes with decreasing volume,
favoring larger bulk modulus for the lighter isotope.

\begin{figure*}[!htb]
	\centering
		\includegraphics[clip=true, trim=20mm 20mm -30mm -20mm, width=1\textwidth]{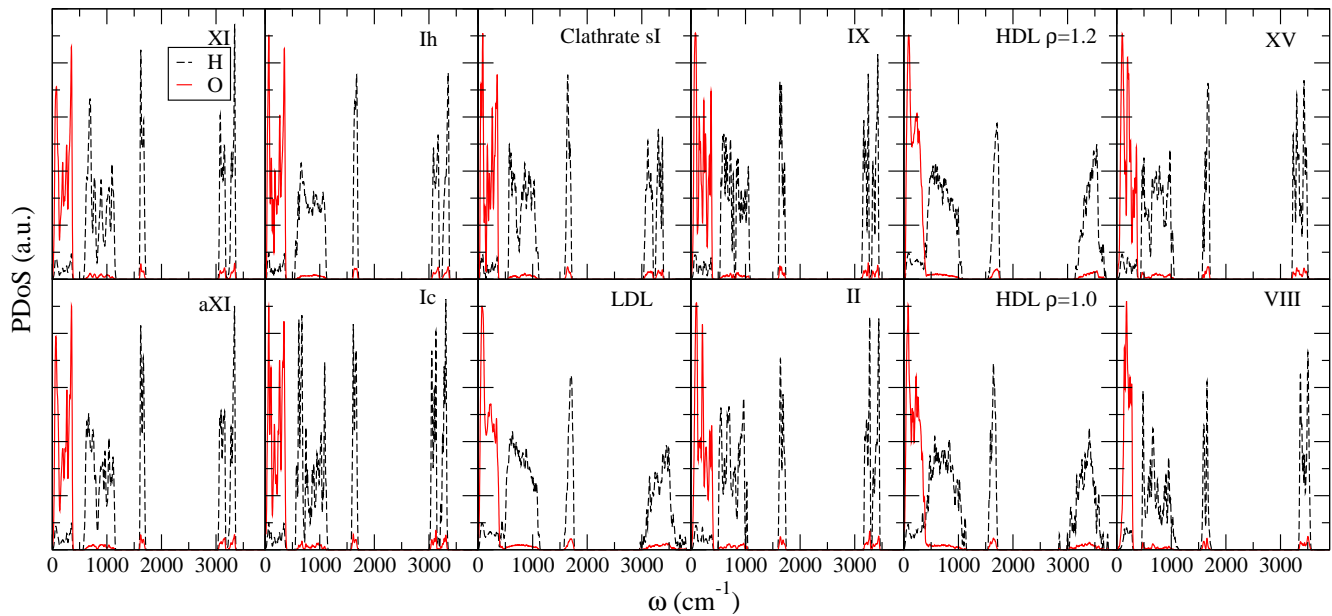}%left bottom right top
	\caption{Phonon density of states for H$_2$O, projected onto H and O atoms,
				obtained with the vdW-DF$^{\rm{PBE}}$ functional.}
	\label{fig:dos1}
\end{figure*}

\begin{figure*}[!htb]
        \centering
	\includegraphics[clip=true, trim=0mm 15mm 0mm 0mm, width=1\textwidth]{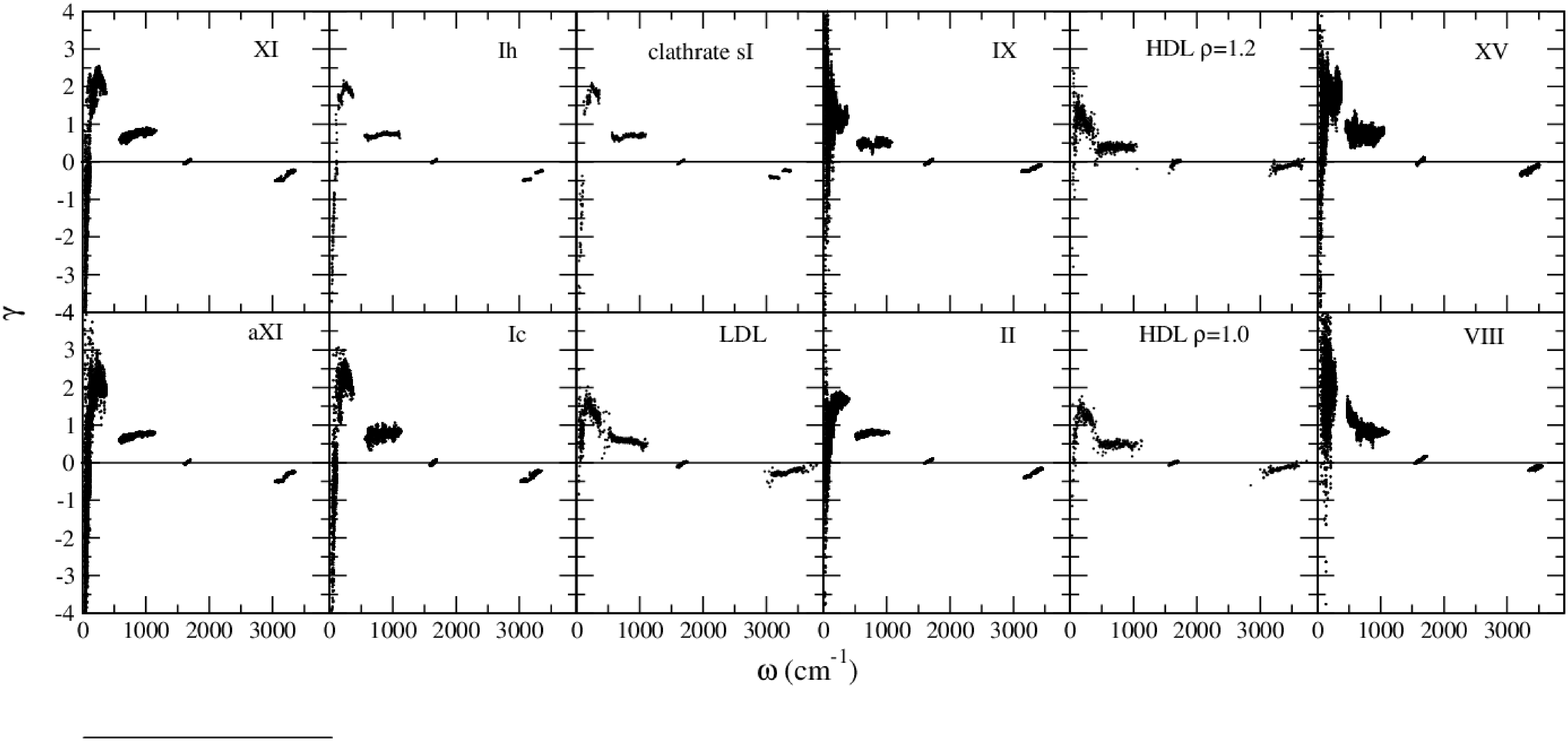}%left bottom right top
        \caption{Average Gr\"uneisen parameter of each phonon mode, obtained with the vdW-DF$^{\rm{PBE}}$ functional.}
        \label{fig:Grun}
\end{figure*}

Within each structure, considering how frequencies and Gr\"uneisen parameters are distributed,
we separate phonon modes into 6 groups.
Fig. \ref{fig:dos1} and \ref{fig:Grun} show phonon density of states and Gr\"uneisen parameters of each mode.
The average phonon frequencies and the corresponding average $\gamma_k$
for each group can be found in the SI of Ref. \cite{Pamuk2015}.
First, there is a band with small frequencies ($\omega\sim74$ cm$^{-1}$), 
associated with the stretching modes of the Hbond, that has negative $\gamma_k$,
responsible for the negative thermal expansion of hexagonal ice. \cite{Kuhs94}
The translational modes ($\omega\sim271$ cm$^{-1}$) are mostly dominated by oxygen atoms, 
while the libration modes ($\omega\sim847$ cm$^{-1}$) are dominated by the hydrogen atoms.
These two modes both contribute to the strength of Hbonding network,
and they have positive $\gamma_k$, favoring the normal isotope effect.
The bending modes ($\omega\sim1646$ cm$^{-1}$) are very harmonic and have very small $\gamma_k$,
and a small effect on the isotope effect.
These modes do not contribute to the interplay between covalent bond and Hbond network.
The highest frequencies correspond to the anti-symmetric ($\omega\sim3314$ cm$^{-1}$ for H$_2$O)
and symmetric ($\omega\sim3115$ cm$^{-1}$ for H$_2$O) stretching modes 
of the OH covalent bond, with weight mostly on the hydrogen atoms.
The stretching modes have negative $ \gamma_k $'s, favoring anomalous isotope effect.
There is an anti-correlation between the translational and librational modes which determine
the strength of the Hbond and the stretching modes which determine the strength of the OH covalent bond.
The anomalous isotope effect is a result of a fine balance between
cancellations of contributions from each band \cite{Pamuk12, Sriram2013}.

Results for the bulk modulus at $T=0$~K are in Table \ref{table:Bulk}.
The temperature dependence of the bulk modulus is given in Fig. \ref{fig:B_T} 
for ferroelectric-proton-ordered ice XI, calculated with the vdW-DF$^{\rm{PBE}}$ functional.

The TTM3-F force field model does not correctly predict
anomalous isotope effect at low temperatures.
Temperature dependence of bulk modulus for ice XI and ice Ih with TTM3-F model
are in SI.
With this model, there is a crossing from normal to anomalous isotope effect at $\sim 270$ K,
close to the melting temperature, where the low frequency modes with $\gamma_k>0$ become classical,
and high frequency modes with $\gamma_k<0$ dominate the quantum effects \cite{Sriram2013}.

%%% UNDO COMMENT

DFT calculations reproduce the anomalous effect on the isotopes of hydrogen atoms, 
i.e the bulk modulus of H$_2$O is larger than the bulk modulus of D$_2$O.
Similarly to the volume isotope effect \cite{Pamuk12}, the bulk moduli of oxygen isotopes are predicted to have a normal isotope effect, 
i.e the bulk modulus of H$_2$O is smaller than the bulk modulus of H$_2\text{}^{18}$O. 
This behavior is reproduced qualitatively for both DFT functionals.
Comparing with previous DFT plane wave studies \cite{Galli12}, 
our PBE calculations perform quite similarly.
However, plane wave calculations that include van der Waals forces with the vdW-DF2 
functional \cite{Lee2010} do not give the anomalous isotope shift.
Therefore, the vdW-DF$^{\rm{PBE}}$ functional is a better candidate for this type of analysis.
There is still room for improvement of van der Waals forces
representation with the density functionals \cite{Santra2013}.

As the temperature increases, the bulk modulus of all ices decrease.
The convergence to the classical bulk modulus occurs at the temperatures much higher than the melting temperature of ice.
This is an indication that the nuclear quantum effects are still important in the liquid water and
must be considered for the correct structural analysis.

\subsection{Nuclear Quantum Effects in Other Ice Polymorphs}

Experiment shows that the anomalous volume isotope effect of the replacement of H with D,
also persists in liquid water, \cite{Kell67, Kell77}
although the structure of the liquid is very different from hexagonal ice.
Therefore, we turn our focus to other crystalline ice phases and ice-like structures with varying densities
in order to understand the isotope effect on the volumes and compressibility.

As before, we start with an analysis of the density of states and the Gr\"uneisen parameters, $\gamma_k$, of all structures,
as given in Fig. \ref{fig:dos1} and \ref{fig:Grun}.
The trends in frequencies and corresponding Gr\"uneisen parameters 
seen in hexagonal ices 
persist in ice Ic and clathrate sI, and also in LDL-like amorphous ice.
However, as the density increases, some of the low frequency bands start mixing and are harder to differentiate into groups.
The first difference appears in the low frequency group with negative $\gamma_k$,
associated with the stretching of the Hbond and the negative Gr\"uneisen parameters and
responsible for the negative thermal expansion of hexagonal ice. \cite{Kuhs94}
This band has negative $\gamma_k$ for ice Ic and clathrate sI;
its Gr\"uneisen parameters start to become positive in amorphous ices, and high density crystalline ices,
ice IX, ice II, ice XV, and ice VIII.
Consequently, a negative thermal expansion is observed in ice Ic at $\sim60$ K and in empty clathrate sI at $\sim70$ K,
and this behavior is not observed in rest of the crystalline phases.
However, this band is not dominant in the determination of the isotope effect in general. 
The translational modes and the libration modes that are linked to the Hbonding network
both mostly have positive $\gamma_k$, favoring the normal isotope effect.
For all structures, the bending modes are very harmonic and they have very small $\gamma_k$,
hence these modes are not effective enough on the isotope effect.
The highest frequencies, which correspond to the anti-symmetric and
symmetric stretching modes of the OH covalent bond with a weight mostly on the hydrogen atoms,
have negative $ \gamma_k $, favoring the anomalous isotope effect.
For the crystalline phases and empty clathrate sI, it is possible to distinguish between the symmetric and anti-symmetric
stretching modes, while for amorphous ices these modes are mixed, similar to the phonon density of states of liquid water.
The anti-symmetric stretching modes have in general less negative $\gamma_k$ as compared to the symmetric
stretching modes.
In addition, these modes become less and less negative as the density of the ice phase increases
going from ice XI towards ice VIII.
Again, the fine balance between the translational and librational modes, which are linked to the strength of the Hbond, 
and the stretching modes, which are linked to the strength of the OH covalent bond,
determine whether the system has a normal or anomalous isotope effect.

\begin{figure}[!htb]
        \centering
                \includegraphics[clip=true, trim=-5mm 10mm 0mm -15mm, width=0.45\textwidth]{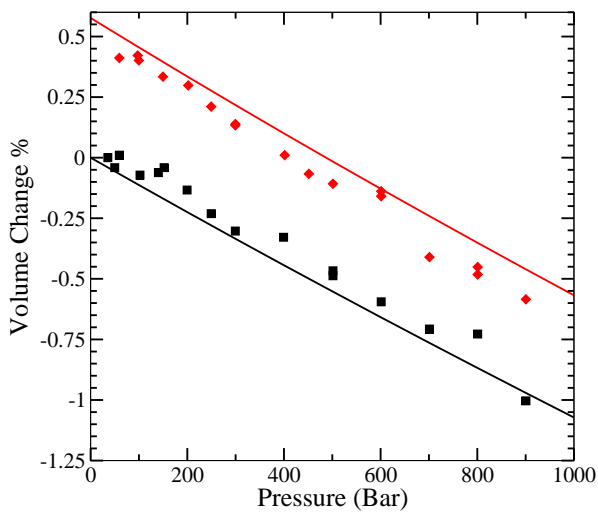}%left bottom right top
        \caption{Volume change $V(P)/V_{\rm{H_2O}}(0)-1$, relative to H$_2$O at $P=0$ bar for empty clathrate structure I.
                 The lines are the calculations and the dots are the experimental values taken from Ref. \cite{Kuhs03}.}
        \label{fig:V_P}
\end{figure}

The isotope effect on the volume of the clathrate sI is known to be anomalous from the experiments \cite{Kuhs03},
therefore we first focus on this structure to compare our calculations.
We calculate the free energy profiles of corresponding experimental temperatures, T = 271 K for H$_2$O and T = 273 K for D$_2$O
and obtain the relation between the pressure and the volume, as explained in eq. (\ref{eq:pres}).
To compare the isotope shift in clathrate sI, the experimental H$_2$O volume at the lowest experimental pressure,
and the calculated H$_2$O volume at zero pressure are both set to zero;
and Fig. \ref{fig:V_P} shows the relative change in the volume from H$_2$O to D$_2$O.
These results show that the anomalous isotope effect in clathrate sI, as well as the amount of the shift in the volume, 
is well captured with the vdW-DF$^{\rm{PBE}}$ functional.
Even though there is a difference in the slopes of the experimental and the theoretical values, which is reflected
in the bulk modulus results, the net isotope shift in the volume is well represented within the QHA. 
The isotope shift is similar for the filled structures, and is presented in the SI.
Similarly to the isotope effect in volume, the isotope effect in the bulk modulus is also anomalous.

\begin{figure*}[!ht]
	\centering
		\includegraphics[clip=true, trim=10mm 10mm -10mm -20mm, width=1\textwidth]{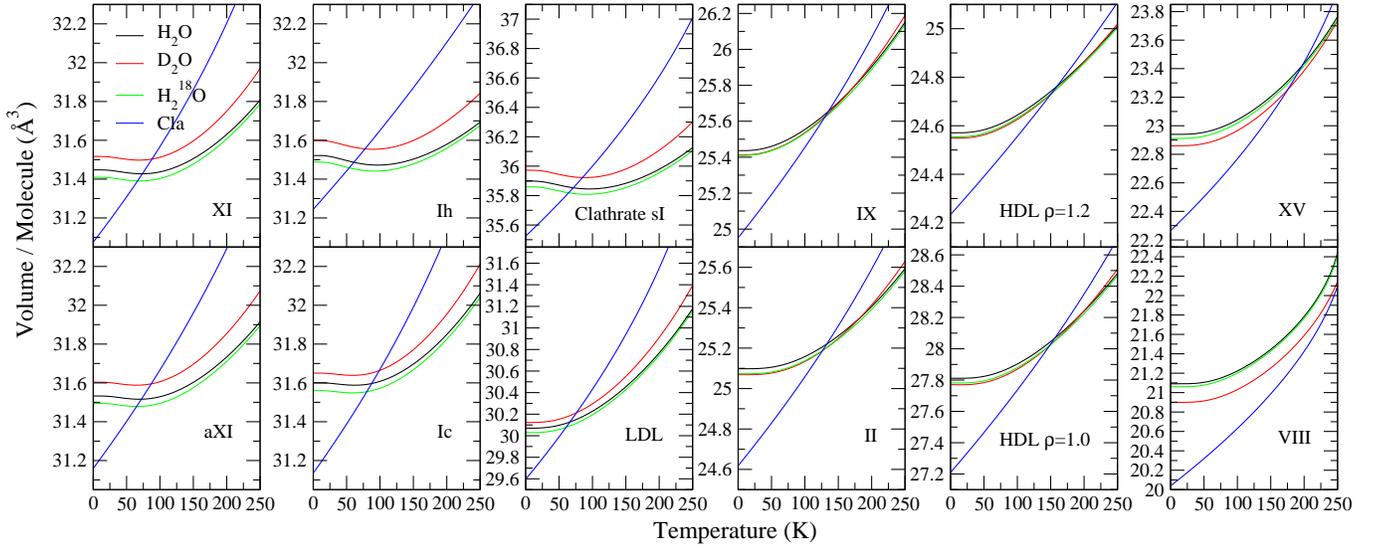}%left bottom right top
	\caption{Volume per molecule as a function of temperature calculated with the QHA using the
	vdW-DF$^{\rm{PBE}}$ functional.}
	\label{fig:vol_T}
\end{figure*}

\begin{figure*}[htb]
	\centering
		\includegraphics[clip=true, trim=60mm 15mm -70mm -20mm, width=1\textwidth]{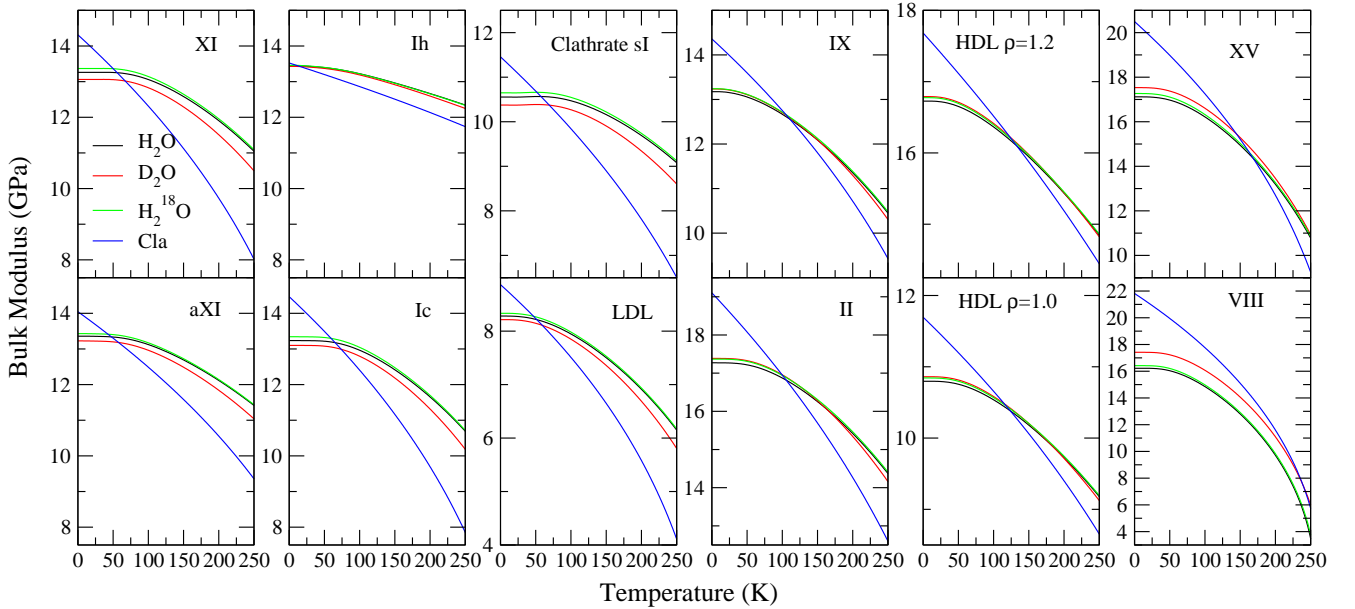}%left bottom right top
	\caption{Bulk modulus as a function of temperature calculated with the QHA using the
	vdW-DF$^{\rm{PBE}}$ functional.}
	\label{fig:bulk_T}
\end{figure*}

Next, we present the volume and bulk modulus as a function of temperature for all structures
in Fig. \ref{fig:vol_T} and Fig. \ref{fig:bulk_T} respectively.
The low density crystalline ices: hexagonal ices and cubic ice have a similar structure with
a volume per molecule of $V\sim 31.5$ \AA~ at $T=0$ K, 
and a bulk modulus of $B\sim 13$ GPa.
The empty clathrate sI has slightly larger volume per molecule $V \sim 36$ \AA, and smaller bulk modulus $B \sim 10.5$ GPa,
and these results are in general good agreement with the experiments \cite{Kuhs03}.
The LDL-like amorphous ice has smaller volume per molecule $V \sim 30$ \AA~ and bulk modulus $B \sim 8$ GPa, which is close to that of liquid water.
The crystalline ices that are in the middle of the pressure range of the phase diagram: ice IX and ice II
both have a volume per molecule of $V \sim 25$ \AA,
and this is in good agreement with the experiments \cite{Londono1993,Kamb1964}.
At $T=225$ K, the volume for ice II is in good comparison with the experimental value of $V=25.6$ \AA,
while the bulk modulus $B \sim 15$ GPa is larger than the experimental value of $B=12.13$ GPa \cite{Fortes2005}.
There is no data, to our knowledge, on the bulk modulus of ice IX,
therefore we comment on its proton-disordered polymorph ice III
that has an experimental bulk modulus very similar to the bulk modulus of ice Ih $B \sim 9.6$ GPa \cite{Gagnon1990},
which agrees with our results that the bulk modulus stays $B \sim 13$ GPa slightly decreasing from ice XI to ice IX.
The volume per molecule decreases as we get to ice XV;
the experimental volume per molecule is $V=22.5$ \AA \cite{Finney2009},
and is slightly smaller than our calculated result of $V\sim23$ \AA.
Again due to the lack of bulk modulus data of ice XV, we compare to its proton-disordered polymorph ice VI,
and the bulk modulus increases significantly to $B=18.5$ GPa \cite{Gagnon88},
which agrees with the trend of the increase in our calculated bulk modulus results.
This trend of decrease in the volume and increase in the bulk modulus also occurs going from LDL-like amorphous ices
to HDL-like amorphous ices.
Ice VIII has the smallest volume per molecule among all the structures, and the calculated value $V\sim21$ \AA~ 
slightly overestimates as compared to experiments under pressure \cite{Kuhs84}.
Consequently, the calculated bulk modulus $B\sim16$ GPa is underestimated as compared to the
recently reported experimental value of $B=18.7$ GPa \cite{Klotz2017},
and also as compared to its proton-disordered polymorph ice VII $B=20.4$ GPa \cite{Adriaan2017,Klotz1994}.
We continue our analysis with the isotope effect, and we present, for completeness, the isotope effect of the oxygen atoms 
both on the volume and bulk modulus, which stays as normal isotope effect for all structures,
but in the following discussion, we focus on the isotope effect of the hydrogen atoms.

\begin{figure}[!htb]
        \centering
                \includegraphics[clip=true, trim=0mm 0mm 0mm 0mm, width=0.45\textwidth]{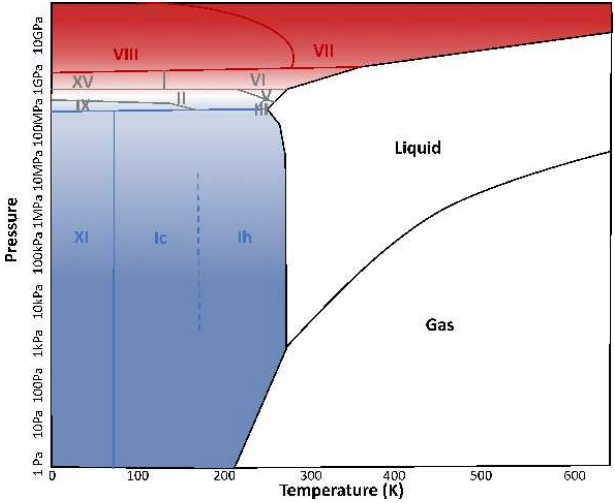}%left bottom right top
        \caption{Isotope effect on the phase diagram of water (adapted from Ref. \onlinecite{phase}).
        Blue, low pressure region: ice phases with the anomalous isotope effect. 
        White, crossing region: ice phases at the crossing of normal to anomalous isotope effect. 
        Red, high pressure region: ice phases with the normal isotope effect.
		}
        \label{fig:phase}
\end{figure}

The isotope effect at $T=0$ K is anomalous and it stays anomalous up to the melting point
for low density structures: hexagonal ices, ice Ic, clathrate sI and LDL-like ices.
As the density of the structure increases, going towards ice IX, ice II, and ice XV, as well as in HDL-like amorphous ices, 
the isotope effect is normal at low temperatures, with a transition to the
anomalous isotope effect at higher temperatures.
Although it is not possible to exactly determine the temperature of the crossing from normal isotope effect to 
anomalous isotope effect, we can observe qualitatively that the higher the density, the higher the crossing temperatures.
And finally, the isotope effect is normal up to the melting point for ice VIII, which has the highest density.
All the results are given at zero pressure, however some of these structures are stable only under applied pressure.
Therefore, we also present the isotope effect in volume of ice VIII under 2.5 GPa as a comparison in the SI,
and the conclusions we draw do not change.
The trends on the isotope effect on the bulk modulus follow the isotope effect on the volume.
The isotope effect on the bulk modulus also starts anomalous at low density ice structures 
and then presents a crossing from normal to anomalous isotope effect
as the density of the structure increases.
Finally, at the highest density structure of ice VIII, it stays normal up to the melting point.
This is depicted on the phase diagram in Fig. \ref{fig:phase} as adapted from Ref. \onlinecite{phase}.
Blue shaded area shows the ice phases that have the anomalous effect, the crossing region in white shows the ice phases
that are at the crossing of the normal and anomalous effects, and the red region shows the ice phases that have the normal isotope effect.

\begin{table*} [ht] \footnotesize
\caption{Average oxygen-oxygen distance in \AA~ for Hbonded and vdW bonded configurations and
              average number of bonds per molecule for different configurations of structures.
             The quantum H$_2$O volume in \AA$^3$ and bulk modulus in GPa with their isotope effect
        		from hydrogen to deuterium, IS$_V$(H-D)=$\frac{V({\rm H})}{V({\rm D})}-1$ and IS$_B$(H-D)=$\frac{B({\rm H})}{B({\rm D})}-1$ 
        		at zero temperature and pressure are also included.}
\centering
\begin{ruledtabular}
%\scalebox{1.0}{
                \begin{tabular} {l c c c c c c r l}
%                \hline
%                \hline
%                \cline{1-9}
                Ice & $<r_{\rm OO}^{\rm{Hbond}}>$ & $<N^{\rm{Hbond}}>$ & $<r_{\rm OO}^{\rm{vdW}}>$ & $<N^{\rm{vdW}}>$ & $V_{\rm{H_2O}}$ & IS$_V$(H-D) & $B_{\rm{H_2O}}$ & IS$_B$(H-D) \\
                \hline
                XI                             & 2.719 & 4.000 & 0.000 & 0.000 & 31.41 & $-0.32\%$ & 13.26 & $+1.51\%$ \\
                Clathrate sI                   & 2.728 & 4.000 & 0.000 & 0.000 & 35.90 & $-0.20\%$ & 10.56 & $+1.71\%$ \\
                LDL-like$^{\rho=1.0}_{T=300}$  & 2.765 & 3.953 & 3.899 & 0.688 & 30.07 & $-0.18\%$ &  8.28 & $+0.81\%$ \\
                Ic                             & 2.719 & 4.000 & 0.000 & 0.000 & 31.60 & $-0.16\%$ & 13.24 & $+1.02\%$ \\
                HDL-like$^{\rho=1.2}_{T=300}$  & 2.815 & 4.047 & 3.582 & 2.109 & 24.57 & $+0.09\%$ & 16.73 & $-0.38\%$ \\
                IX                             & 2.737 & 4.000 & 3.489 & 2.000 & 25.44 & $+0.10\%$ & 13.18 & $-0.456\%$ \\
                II                             & 2.756 & 4.000 & 3.487 & 2.000 & 25.10 & $+0.12\%$ & 17.27 & $-0.666\%$ \\
                HDL-like2$^{\rho=1.0}_{T=300}$ & 2.783 & 4.063 & 3.748 & 0.969 & 27.81 & $+0.15\%$ & 10.80 & $-0.59\%$ \\
                HDL-like1$^{\rho=1.0}_{T=300}$ & 2.790 & 4.063 & 3.622 & 0.969 & 27.71 & $+0.28\%$ &  9.90 & $-1.66\%$ \\
                HDL-like$^{\rho=1.0}_{T=350}$  & 2.917 & 3.781 & 3.834 & 1.938 & 27.63 & $+0.35\%$ &  9.69 & $-5.21\%$ \\
		XV			       & 2.773 & 4.000 & 3.354 & 2.800 & 22.94 & $+0.35\%$ & 17.12 & $-2.40\%$ \\
                VIII                           & 2.943 & 4.000 & 2.949 & 4.000 & 21.09 & $+0.90\%$ & 16.22 & $-7.40\%$ \\
%                \hline
%                \hline
                \end{tabular}
%                }
\end{ruledtabular}
\label{table:avebonds}
\end{table*}

These observations let us make one final analysis on these systems.
Table \ref{table:avebonds} shows the average distance between oxygen atoms when they are
Hbonded, $<r_{\rm OO}^{\rm{Hbond}}>$, van der Waals bonded $<r_{\rm OO}^{\rm{vdW}}>$,
and the average number of each bond in the analyzed configurations.
Also included are the isotope effect with respect to H$_2$O and its quantum volume.
The structures are listed from the most anomalous to the least anomalous in the isotope effect on the volume.

For the structures with anomalous effect, the Hbond length is small and is highly populated,
while those systems have no (ice XI, clathrate sI, ice Ic) or very few (LDL-like amorphous) vdW bonds.
As the density of the structures increases (ice IX, HDL-like amorphous, ice II),
Hbond length keeps increasing, but more importantly, these systems have more van der Waals bonds,
populating the interstitial sites.
Then the effect starts to become normal, but small,
and a transition to anomalous isotope effect happens at high temperatures.
In the most extreme case of ice VIII, both Hbond length is the largest, vdW bond length is the smallest
and the number of vdW bonds is the largest.
Then, for all temperatures and pressures, the isotope effect stays normal.

This shows that the density or the $r_{\rm OO}$ distance by itself is not enough to predict the sign
of the isotope effect.
For example for HDL-like$^{\rho=1.2}_{T=300}$ structure,
in spite of the fact that $r_{\rm OO}$ distance of both bonds are larger than ice IX and ice II,
the volume is smaller (density is larger).
This is because both bonds are more populated, leading to a conclusion of smaller volume per molecule
and a normal isotope effect, after all.
Another example is ice VIII where the experimental Hbond distance is
$r_{\rm{OO}}^{\rm{Hbond}}=2.979(1)$\AA~ \cite{Kuhs84},
and for the vdW-DF$^{\rm{PBE}}$ functional it is $r_{\rm{OO}}^{\rm{Hbond}}=2.943$\AA.
Then, each molecule has four vdW bonded (non-bonded) configurations, two of which are closer than the Hbond distance, and two  further away.
Experimentally, the vdW bonded configuration distance is: $r_{\rm{OO}}^{\rm{vdW1}}=2.743(9)$\AA~ \cite{Kuhs84} and
for vdW-DF$^{\rm{PBE}}$ functional it is $r_{\rm{OO}}^{\rm{vdW1}}=2.822$\AA.
The second vdW bonded configuration distance for vdW-DF$^{\rm{PBE}}$ functional is: $r_{\rm{OO}}^{\rm{vdW2}}=3.075$\AA,
resulting an average of 4 vdW bond lengths of $r_{\rm{OO}}^{\rm{vdW}}=2.949$\AA.
Therefore, there is a fine balance between the bond length distances and how many bonds
each molecule has.
This discussion agrees with recent predictions on the isotope effect in other ice phases \cite{Wentzcovitch2015, Wentzcovitch2017}.

\begin{figure}[!htb]
        \centering
                \includegraphics[clip=true, trim=0mm 10mm 0mm -15mm, width=0.45\textwidth]{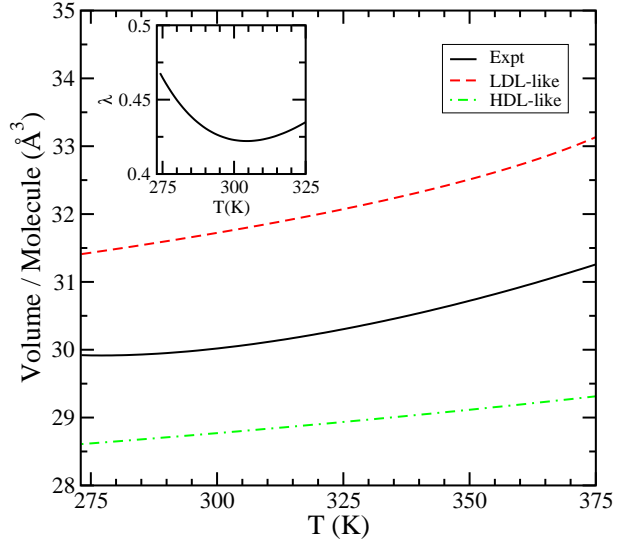}%left bottom right top
        \caption{Volume per molecule calculated with the QHA at the liquid water temperatures
        for the LDL-like amorphous ice and the HDL-like amorphous ice is compared to the experimental values for the liquid.
        The inset shows what the LDL-like contribution $\lambda$, such that $\lambda V^{\rm LDL-like}+(1-\lambda)V^{\rm HDL-like}=V^{\rm expt.}$
		}
        \label{fig:lambda}
\end{figure}

Finally, we can make some links to liquid water, by exploiting the experimental
results on the isotope effect on the volume of liquid water \cite{Kell67, Kell77}.
Considering the difference between the volume per molecule
of liquid water with D$_2$O and H$_2$O,
$V_{\rm D_2O}-V_{\rm H_2O}$, from Fig. \ref{fig:waterV_T}, the volume difference decreases with increasing temperature,
therefore the isotope effect goes from strongly anomalous towards weakly anomalous,
but not crossing up to the boiling temperature.
There are two different regions: 
a region with rapidly decreasing difference 
$V_{\rm D_2O}-V_{\rm H_2O}$
going from anomalous towards a normal isotope effect, up to $T \sim 310$ K, 
and a second region with almost constant, slowly decreasing difference, up to the melting temperature.
As the Hbonding network is never completely broken in liquid water, as in ice VIII, 
the isotope effect stays anomalous throughout all the temperatures.
Drawing conclusions from our results, we speculate that the anomalous isotope effect in liquid water
goes from strongly anomalous to weakly anomalous as the temperature increases from the melting point
to the bulk modulus maximum (compressibility minimum), 
therefore the LDA-like components of the liquid decrease in this region.
Then it starts behaving as a normal liquid after this temperature.

This is also shown in Fig. \ref{fig:lambda}, where the volume per molecule of the liquid water from Ref. \onlinecite{Kell77}
is compared to one of the configurations with the LDL-like amorphous ice and to one of the configurations with the HDL-like amorphous ice,
as calculated with the QHA at high temperatures.
Even though the quantitative values are not exact at these temperatures with the QHA,
one can estimate the LDL-like and HDL-like contributions to the liquid by approximating
the liquid water volume per molecule such that $V^{\rm expt.}=\lambda V^{\rm LDL-like}+(1-\lambda)V^{\rm HDL-like}$.
Then it is clear that up to $\sim 305$ K, the LDL-like contributions decrease.
Therefore, a force-field model that can reproduce the aforementioned anomalies of water must also reproduce the
anomalies in the isotope effect.

\section{Conclusion}

We study the nuclear quantum effects on the volume and bulk modulus of several ice phases, 
as well as ice-like structures.
We start with a structural analysis of the lattice parameters and the volume of hexagonal ices,
including different components of the strain tensor
and we show that the classical (frozen) bulk modulus along the x-y plane is smaller than along the z-axis,
hence the hexagonal ice is softer along the x-y plane.
In addition, we calculate the isotope effect on the total bulk modulus of the hexagonal ices.
There is an anomalous isotope effect on the bulk modulus for hydrogen isotopes:
the bulk modulus of D$_2$O is smaller than H$_2$O,
and normal isotope effect on the bulk modulus for oxygen isotopes;
similarly to the isotope effect on the volume.

Then we investigate the isotope effect on different polymorphs of ice.
First, we obtain an anomalous isotope effect on the volume of the clathrate structure I,
in very good agreement with the experiments.
Then we calculate that the isotope effect is anomalous in ice Ih, clathrate sI, ice Ic and LDL-like amorphous ices.
Moving on to high pressure ice phases, where the interstitial sites start to get filled,
in the intermediate ice phases, ice IX, ice II, and ice XV, as well as in the HDL-like amorphous ices,
we observe that the difference between the volumes of H and D
starts as a small and normal isotope effect and becomes anomalous at
higher temperatures with the increasing density.
Once we get to the other end of the phase diagram, ice VIII is the most dense ice phase,
where the interstitial sites are completely filled, the isotope effect
becomes completely normal both at zero pressure and at high pressures.

Finally, we discuss the implications of our results on the isotope effect in liquid water.
When the hexagonal ice Ih, which is already a structure with large anomalous isotope effect, melts,
the effect in liquid water also starts as large and anomalous.
With the increasing temperature, the anomaly of the isotope effect (difference in the volume of the two isotopes) 
decreases rapidly up to $T \sim 310$ K, which is close to isothermal compressibility minimum temperature, $T \sim 320$ K.
Until the compressibility minimum temperature, as the interstitial sites are being filled,
the isotope effect goes towards being normal very quickly.
This is similar to moving from ice Ih towards ice VIII in the analyzed structures.
However, in liquid water, the Hbond network is intact as compared to ice VIII.
Therefore, the net effect stays anomalous.
Once the compressibility minimum is reached, then the change in the isotope effect is
stabilized and it keeps on decreasing at a much slower rate, until it gets to the boiling temperature.
With this, we can speculate that the nuclear quantum effects and the anomalous isotope effect
are linked to the compressibility minimum of liquid water.
This conclusion opens new horizons for the development of semi-empirical water models,
showing that a complete theory should be able to explain different anomalies of water,
in different regions of the phase diagram.

\section{Acknowledgments}

This work is supported by DOE Grants No. DE-SC0003871 (M.V.F.S.), and No. DE- FG02-08ER46550 (P.B.A.) 
and the grant FIS2012-37549-C05 from the Spanish Ministry of Economy and Competitiveness. 
This work used the resources 
at the Center for Functional Nanomaterials, Brookhaven National Laboratory, 
	which is supported by the U.S. Department of Energy, Office of Basic Energy Sciences, under Contract No. DE-AC02-98CH10886
	and No. DE-SC0012704;
at the Extreme Science and Engineering Discovery Environment (XSEDE), 
	which is supported by National Science Foundation grant number ACI-1548562;
and at the Handy and LI-Red computer clusters at the Stony Brook University Institute for Advanced Computational Science.
B.P. acknowledges National Science Foundation [Platform for the Accelerated Realization, Analysis, and Discovery of Interface Materials (PARADIM)] under Cooperative Agreement No. DMR-1539918 for her time at Cornell University.

\section{Supporting Information}

In the supporting information, we present a detailed derivation of the relations between the strain tensor
and the bulk modulus, that is used to calculate the anisotropy in the bulk modulus of the hexagonal ices.
We also present the temperature dependence of the bulk modulus of the hexagonal ices 
as calculated with the TTM3-F force-field model,
comparison of the isotope effect of the filled and empty clathrate structures, 
and the isotope effect on the volume of ice VIII at $P=2.5$ GPa.

\subsection{Strain Tensor and Bulk Modulus Relations}
                                                                                                                                                                       
When there is a small uniform deformation on a solid, the axes are distorted in orientation by $\epsilon_{\alpha\beta}$. \cite{Kittel}
Then the displacement of an atom due to this deformation can be defined as:

\begin{eqnarray}
\mathbf{R}(\mathbf{r})&\equiv&(x\epsilon_{xx}+y\epsilon_{yx}+z\epsilon_{zx})\hat{x}+(x\epsilon_{xy}+y\epsilon_{yy}+z\epsilon_{zy})\hat{y} \nonumber \\
					  &&+(x\epsilon_{xz}+y\epsilon_{yz}+z\epsilon_{zz})\hat{z} \nonumber \\
                      &=&u(\mathbf{r})\hat{x}+v(\mathbf{r})\hat{y}+w(\mathbf{r})\hat{z}.
\end{eqnarray}

From the displacement, the coefficients of the strain tensor can be defined by the relation
\begin{equation}
e_{\alpha\beta}\equiv\epsilon_{\alpha\beta}=\frac{\partial u_\alpha}{\partial x_\beta} + \frac{\partial u_\beta} {\partial x_\alpha},
\end{equation} 
where $\alpha$ and $\beta$ runs over $\hat{x}, \hat{y}, \hat{z}$ directions.

Within the approximation of Hooke's law, we can write the elastic energy density as a quadratic function of the strains as follows,
\begin{equation}
\frac{\Delta E}{V}=\frac{1}{2} \sum_{\lambda=1}^6 \sum_{\mu=1}^6 \tilde{C}_{\lambda\mu}e_{\lambda}e_{\mu},
\label{eq:deltaEgeneral}
\end{equation}
where $1\equiv xx$; $2\equiv yy$; $3\equiv zz$; $4\equiv yz$; $5\equiv zx$; $6\equiv xy$.

Noting that only certain combinations enter the stress-strain relations, the elastic stiffness constants are symmetrical:
\begin{equation}
C_{\alpha\beta}=\frac{1}{2}\left(\tilde{C}_{\alpha\beta}+\tilde{C}_{\beta\alpha}\right)=C_{\beta\alpha}.
\end{equation}

Furthermore, we change the lattice parameters along the lattice directions in the hexagonal ice system.
Therefore, we are interested only in symmetry-preserving strains: $e_{xx}=e_{yy}=\Delta a/a$ and $e_{zz}=\Delta c/c$ and there are no shear strains.
This reduces the elastic stiffness constants to the upper left $3\times 3$ part of the elasticity tensor.
Then eq. (\ref{eq:deltaEgeneral}) becomes;
\begin{eqnarray}
\frac{\Delta E}{V}&=&\frac{1}{2}[\tilde{C}_{11}e_1e_1+\tilde{C}_{22}e_2e_2+\tilde{C}_{33}e_3e_3  \nonumber \\
					&& +(\tilde{C}_{12}+\tilde{C}_{21})e_1e_2 \nonumber \\
					&& +(\tilde{C}_{13}+\tilde{C}_{31})e_1e_3 \nonumber \\
					&& +(\tilde{C}_{23}+\tilde{C}_{32})e_2e_3] \nonumber \\
				  &=&\frac{1}{2}[C_{11}e_{xx}^2+C_{22}e_{yy}^2+C_{33}e_{zz}^2 + 2C_{12}e_{xx}e_{yy}  \nonumber \\
				    && + 2C_{13}e_{xx}e_{zz} +2C_{23}e_{yy}e_{zz}] \nonumber \\
				  &=&\frac{1}{2}[2C_{11}e_{xx}^2+C_{33}e_{zz}^2+2C_{12}e_{xx}^2+ 4C_{13}e_{xx}e_{zz}] \nonumber \\
				  &=&\frac{1}{2} \left[ 2(C_{11}+C_{12})\left(\frac{\Delta a}{a}\right)^2 +C_{33}\left(\frac{\Delta c}{c}\right)^2 \right. \nonumber \\
				    &&+\left. 4C_{13}\left(\frac{\Delta a}{a}\right)\left(\frac{\Delta c}{c}\right)\right].
\label{eq:deltaE}
\end{eqnarray}

More simply, we can also write this as:
\begin{equation}
\resizebox{1.5\hsize}{!}{$
\frac{\Delta E}{V} = \frac{1}{2} \left( \begin{array}{ccc}\Delta a/a, & \Delta a/a, & \Delta c/c \end{array}\right) 
\left( \begin{array}{ccc}c_{11}&c_{12}&c_{13} \\  c_{12}&c_{11}&c_{13} \\  c_{13}&c_{13}&c_{33}\end{array}\right)
\left(\begin{array}{c}\Delta a/a \\ \Delta a/a \\ \Delta c/c \end{array}\right).$
}
%%\label{eq:}
\end{equation}
%\begin{equation}
%\frac{\Delta E}{V} = \frac{1}{2} \left( \begin{array}{ccc}\Delta a/a, & \Delta a/a, & \Delta c/c \end{array}\right) 
%\left( \begin{array}{ccc}c_{11}&c_{12}&c_{13} \\  c_{12}&c_{11}&c_{13} \\  c_{13}&c_{13}&c_{33}\end{array}\right)
%\left(\begin{array}{c}\Delta a/a \\ \Delta a/a \\ \Delta c/c \end{array}\right)
%\end{equation}

Let us now consider dilation under hydrostatic pressure,
\begin{equation}
\delta \equiv \frac{V'-V}{V}=e_{xx}+e_{yy}+e_{zz}.
\end{equation}
By using the definition of bulk modulus, we can link dilation to energy as
$\frac{\Delta E}{V} = \frac{1}{2} B \delta^2$.
Combining this with the last part of eq. (\ref{eq:deltaE}), we can see that when we only change lattice parameter $a$, the dilation is 
$\delta=2\Delta a/a$ and bulk modulus we get corresponds to;
\begin{eqnarray}
&&\frac{\Delta E}{V}=\frac{1}{2}[2(C_{11}+C_{12})\frac{\delta^2}{4}] \nonumber \\
&&B_a=\frac{C_{11}+C_{12}}{2}.
\end{eqnarray}

Similarly, when we change only lattice parameter c, the dilation is
$\delta=\Delta c/c$ and bulk modulus we get corresponds to;
\begin{eqnarray}
&&\frac{\Delta E}{V}=\frac{1}{2}C_{33}\delta^2 \nonumber \\
&&B_c=C_{33}.
\end{eqnarray}
This shows that for hexagonal ice, by varying values of $a$ and $c$ near the minimum, we acquire 
theoretical values of three constants of interest: $(C_{11}+C_{12})$, $C_{13}$, and $C_{33}$.  
There are three other elastic constants: 
$C_{11}-C_{12}$, $C_{14}$ and $C_{44}$ that one can only compute by calculating energies of sheared structures.

To obtain $C_{13}$ let us again consider hydrostatic pressure.
The stress tensor is then $\sigma_{\alpha\beta}=P\delta_{\alpha\beta}$.
In 6-component vector notation, we only need the first three components.  
The relation between stress and symmetry-conserving strain becomes
\begin{equation}
 \left( \begin{array}{c} P \\ P \\ P\end{array}\right) = -
\left( \begin{array}{ccc}C_{11}&C_{12}&C_{13} \\  C_{12}&C_{11}&C_{13} \\  C_{13}&C_{13}&C_{33}\end{array}\right)
\left(\begin{array}{c}\Delta a/a \\ \Delta a/a \\ \Delta c/c \end{array}\right).
%\label{eq:}
\end{equation}
If we equate the pressure at the first two equations from the matrix, we get the relations between $\Delta a/a$ and $\Delta c/c$:

\begin{equation}
(C_{11}+C_{12}-2C_{13})\frac{\Delta a}{a}=(C_{33}-C_{13})\frac{\Delta c}{c}.
\end{equation}

This gives us relations between $\Delta V/V$ and $\Delta a/a$ and $\Delta c/c$,
\begin{eqnarray}
%\resizebox{.9\hsize}{!}{$
\frac{\Delta V}{V} = 2\frac{\Delta a}{a} + \frac{\Delta c}{c} &=& \left[\frac{2(C_{33}-C_{13})}{C_{11}+C_{12}-2C_{13}}+1\right]
\frac{\Delta c}{c} \nonumber \\
&=& \left[ 2+\frac{C_{11}+C_{12}-2C_{13}}{C_{33}-C_{13}}\right]\frac{\Delta a}{a}
%$}
%\label{eq:}
\end{eqnarray}

Finally, the bulk modulus is defined by $P=-B\Delta V/V$, from the second equation of the P matrix
\begin{eqnarray}
B &=& (2C_{13}\frac{\Delta a}{a}+C_{33}\frac{\Delta c}{c}) \nonumber \\
   && (\frac{C_{33}-C_{13}}{2C_{33}-2C_{13}+C_{11}+C_{12}-2C_{13}}\frac{a}{\Delta a}) \nonumber \\
  &=& (2C_{13}\frac{\Delta a}{a}+C_{33}\frac{(C_{11}+C_{12}-2C_{13})}{C_{33}-C_{13}}\frac{\Delta a}{a}) \nonumber \\
   && (\frac{C_{33}-C_{13}}{2C_{33}-4C_{13}+C_{11}+C_{12}}\frac{a}{\Delta a}) \nonumber \\
  &=& \frac{2C_{13}(C_{33}-C_{13})+C_{33}(C_{11}+C_{12}-2C_{13})}{2C_{33}-4C_{13}+C_{11}+C_{12}} \nonumber \\
  &=& \frac{C_{33}(C_{11}+C_{12})-2C_{13}^2}{C_{11}+C_{12}+2C_{33}-4C_{13}}.
\label{eq:bulktot}
\end{eqnarray}

Now that we know $B$, $(C_{11}+C_{12})$, and $C_{33}$, we can use eq. (\ref{eq:bulktot}) to obtain $C_{13}$.

\subsection{TTM3-F Model Temperature Dependence of Bulk Modulus}

\begin{figure}[htb]
        \centering
                \includegraphics[clip=true, trim=0mm 13mm -3mm -13mm, width=0.4\textwidth]{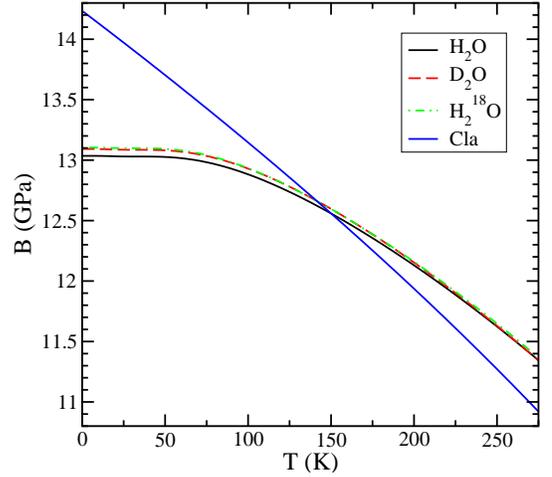}%left bottom right top
        \caption{Bulk modulus as a function of temperature calculated with the TTM3-F force field model for H-ordered ice XI.
                 Note that the anomalous isotope effect is recovered at $\sim$ 270 K.}
        \label{fig:B_T_XI}
\end{figure}

\begin{figure}[htb]
        \centering
                \includegraphics[clip=true, trim=0mm 13mm -3mm -13mm, width=0.4\textwidth]{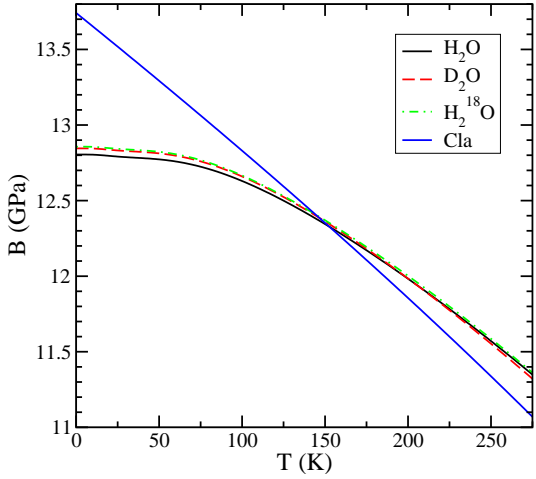}%left bottom right top
        \caption{Bulk modulus as a function of temperature calculated with the TTM3-F force field model for H-disordered ice Ih.
                 Note that the anomalous isotope effect is recovered at $\sim$ 200 K.}
        \label{fig:B_T_Ih}
\end{figure}

It is important to note that TTM3-F model is still capable of reproducing the anomalous isotope effect at high temperatures.
The normal to anomalous isotope effect transition occurs at $\sim$ 270 K for ice XI and at $\sim$ 200 K for ice Ih,
as shown in Fig. \ref{fig:B_T_XI} and \ref{fig:B_T_Ih}, respectively.
At these temperatures, the most stable phase is already ice Ih, therefore, TTM3-F works well with predicting the stable phase
only at high temperatures, close to the melting point.

\subsection{Volume as a Function of Pressure in Clathrate Hydrate Structure I}

\begin{figure}[htb]
        \centering
                \includegraphics[clip=true, trim=-3mm 13mm -3mm -13mm, width=0.45\textwidth]{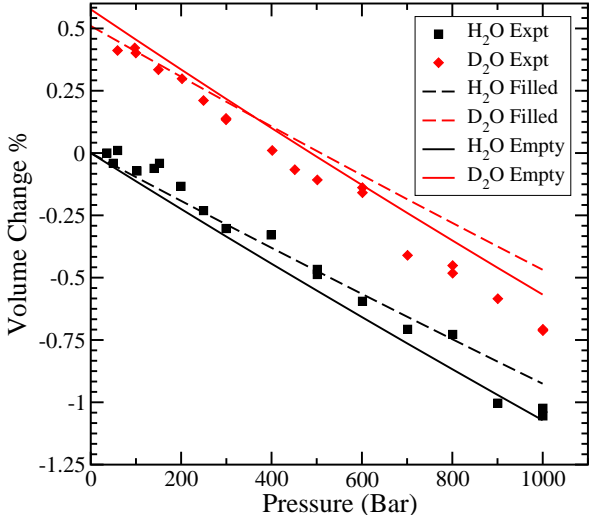}%left bottom right top
        \caption{Volume change $V(P)/V_{\rm H_2O}$ as a function of pressure for filled and empty clathrate hydrates of structure I.
        Note that the isotope effect stays anomalous regardless of the filling of the clathrate sI, and
        the main difference is in the slope of $V(P)$, therefore reflected in the exact value of the corresponding bulk modulus.}
        \label{fig:VPclathr}
\end{figure}

In Fig. \ref{fig:VPclathr}, we present the isotope effect as a volume change \% relative to H$_2$O at $P=0$ bar,
$V(P)/V_{\rm H_2O}$, for the filled clathrate structure as compared to the empty structure presented in Fig. \ref{fig:V_P} %6
in the main paper. 
The isotope effect stays anomalous regardless of the existence of the host molecules, 
and the main difference is in the slopes of the $V(P)$ lines.

\begin{figure}[htb]
        \centering
                \includegraphics[clip=true, trim=-8mm 13mm -3mm -13mm, width=0.45\textwidth]{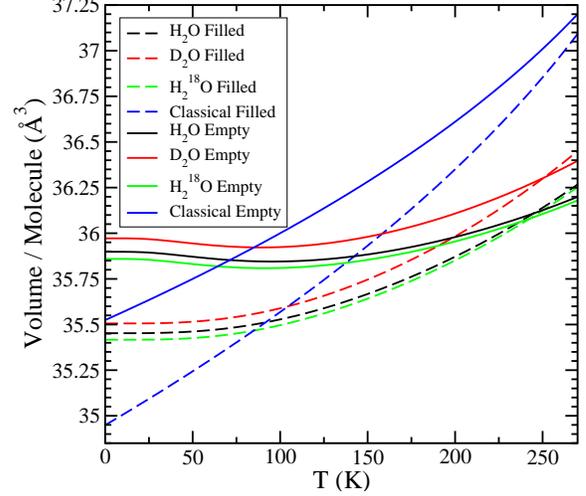}%left bottom right top
        \caption{Volume per molecule as a function of temperature for filled and empty clathrate hydrate structure I.
        The solid lines are the empty structure and the dashed lines are the filled structure.}
        \label{fig:VTclathr}
\end{figure}

We present the empty clathrate sI results in the main paper for a fair comparison with the other ice structures.
Here we also present the results for the clathrate hydrate structure I with 8 CH$_4$ host molecules.

Fig. \ref{fig:VTclathr} shows the volume per molecule as a function of temperature for the filled clathrate structure 
as compared to the empty structure presented in Fig. \ref{fig:vol_T} %7
in the main paper.
Again the isotope effect is anomalous regardless of the host molecules.
We also find that the volume per water molecule of the filled structure is smaller than that of the empty structure,
in agreement with the experimental results on clathrates.

\subsection{Isotope Effect on the Volume of Ice VIII at High Pressure}

\begin{figure}[!htb]
        \centering
                \includegraphics[clip=true, trim=-8mm 13mm -3mm -13mm, width=0.45\textwidth]{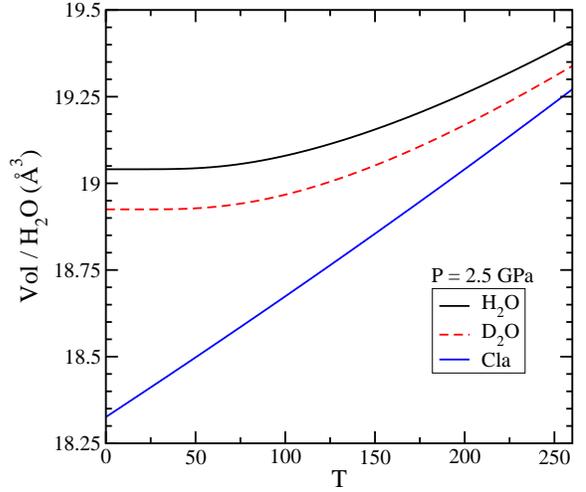}%left bottom right top
        \caption{Volume per molecule as a function of temperature for ice VIII at $P=2.5$ GPa 
        calculated with the QHA using the vdW-DF${\rm PBE}$ functional.}
        \label{fig:V_P2.5}
\end{figure}

Ice VIII is a stable phase under applied pressure. In Fig. \ref{fig:V_P2.5}, we present the volume per molecule as a function of
temperature under applied pressure of $P=2.5$ GPa.

We compare the experimental volume per molecule $V=18.4$ \AA~ at $P=2.4$ GPa \cite{Kuhs84} with the calculated
volume of $V = 19.06$ \AA, and the calculated volume is slightly overestimated.
The isotope effect on ice VIII stays normal up to the melting point regardless of the pressure.

%\clearpage

\bibliography{PaperBetul}

\begin{thebibliography}{86}%
\makeatletter
\providecommand \@ifxundefined [1]{%
 \@ifx{#1\undefined}
}%
\providecommand \@ifnum [1]{%
 \ifnum #1\expandafter \@firstoftwo
 \else \expandafter \@secondoftwo
 \fi
}%
\providecommand \@ifx [1]{%
 \ifx #1\expandafter \@firstoftwo
 \else \expandafter \@secondoftwo
 \fi
}%
\providecommand \natexlab [1]{#1}%
\providecommand \enquote  [1]{``#1''}%
\providecommand \bibnamefont  [1]{#1}%
\providecommand \bibfnamefont [1]{#1}%
\providecommand \citenamefont [1]{#1}%
\providecommand \href@noop [0]{\@secondoftwo}%
\providecommand \href [0]{\begingroup \@sanitize@url \@href}%
\providecommand \@href[1]{\@@startlink{#1}\@@href}%
\providecommand \@@href[1]{\endgroup#1\@@endlink}%
\providecommand \@sanitize@url [0]{\catcode `\\12\catcode `\$12\catcode
  `\&12\catcode `\#12\catcode `\^12\catcode `\_12\catcode `\%12\relax}%
\providecommand \@@startlink[1]{}%
\providecommand \@@endlink[0]{}%
\providecommand \url  [0]{\begingroup\@sanitize@url \@url }%
\providecommand \@url [1]{\endgroup\@href {#1}{\urlprefix }}%
\providecommand \urlprefix  [0]{URL }%
\providecommand \Eprint [0]{\href }%
\providecommand \doibase [0]{http://dx.doi.org/}%
\providecommand \selectlanguage [0]{\@gobble}%
\providecommand \bibinfo  [0]{\@secondoftwo}%
\providecommand \bibfield  [0]{\@secondoftwo}%
\providecommand \translation [1]{[#1]}%
\providecommand \BibitemOpen [0]{}%
\providecommand \bibitemStop [0]{}%
\providecommand \bibitemNoStop [0]{.\EOS\space}%
\providecommand \EOS [0]{\spacefactor3000\relax}%
\providecommand \BibitemShut  [1]{\csname bibitem#1\endcsname}%
\let\auto@bib@innerbib\@empty
%</preamble>
\bibitem [{\citenamefont {Kell}(1967)}]{Kell67}%
  \BibitemOpen
  \bibfield  {author} {\bibinfo {author} {\bibfnamefont {G.~S.}\ \bibnamefont
  {Kell}},\ }\href@noop {} {\bibfield  {journal} {\bibinfo  {journal} {Journal
  of Chemical \& Engineering Data}\ }\textbf {\bibinfo {volume} {12}},\
  \bibinfo {pages} {66} (\bibinfo {year} {1967})}\BibitemShut {NoStop}%
\bibitem [{\citenamefont {Smirnova}\ \emph {et~al.}(2006)\citenamefont
  {Smirnova}, \citenamefont {Bykova}, \citenamefont {Durme},\ and\
  \citenamefont {Mele}}]{Smirnova2006}%
  \BibitemOpen
  \bibfield  {author} {\bibinfo {author} {\bibfnamefont {N.}~\bibnamefont
  {Smirnova}}, \bibinfo {author} {\bibfnamefont {T.}~\bibnamefont {Bykova}},
  \bibinfo {author} {\bibfnamefont {K.~V.}\ \bibnamefont {Durme}}, \ and\
  \bibinfo {author} {\bibfnamefont {B.~V.}\ \bibnamefont {Mele}},\ }\href@noop
  {} {\bibfield  {journal} {\bibinfo  {journal} {J. Chem. Thermodynamics}\
  }\textbf {\bibinfo {volume} {38}},\ \bibinfo {pages} {879} (\bibinfo {year}
  {2006})}\BibitemShut {NoStop}%
\bibitem [{\citenamefont {Angell}\ \emph {et~al.}(1973)\citenamefont {Angell},
  \citenamefont {Shuppert},\ and\ \citenamefont {Tucker}}]{Angell1973}%
  \BibitemOpen
  \bibfield  {author} {\bibinfo {author} {\bibfnamefont {C.~A.}\ \bibnamefont
  {Angell}}, \bibinfo {author} {\bibfnamefont {J.}~\bibnamefont {Shuppert}}, \
  and\ \bibinfo {author} {\bibfnamefont {J.~C.}\ \bibnamefont {Tucker}},\
  }\href@noop {} {\bibfield  {journal} {\bibinfo  {journal} {J. Phys. Chem.}\
  }\textbf {\bibinfo {volume} {77}},\ \bibinfo {pages} {3092} (\bibinfo {year}
  {1973})}\BibitemShut {NoStop}%
\bibitem [{\citenamefont {Kumar}\ \emph {et~al.}(2008)\citenamefont {Kumar},
  \citenamefont {Franzese},\ and\ \citenamefont {Stanley}}]{Kumar2008}%
  \BibitemOpen
  \bibfield  {author} {\bibinfo {author} {\bibfnamefont {P.}~\bibnamefont
  {Kumar}}, \bibinfo {author} {\bibfnamefont {G.}~\bibnamefont {Franzese}}, \
  and\ \bibinfo {author} {\bibfnamefont {H.~E.}\ \bibnamefont {Stanley}},\
  }\href@noop {} {\bibfield  {journal} {\bibinfo  {journal} {J. Phys.: Condens.
  Matter}\ }\textbf {\bibinfo {volume} {20}},\ \bibinfo {pages} {244114}
  (\bibinfo {year} {2008})}\BibitemShut {NoStop}%
\bibitem [{\citenamefont {Mallamace}\ \emph {et~al.}(2013)\citenamefont
  {Mallamace}, \citenamefont {Corsaro},\ and\ \citenamefont
  {Stanley}}]{Mallamace2013}%
  \BibitemOpen
  \bibfield  {author} {\bibinfo {author} {\bibfnamefont {F.}~\bibnamefont
  {Mallamace}}, \bibinfo {author} {\bibfnamefont {C.}~\bibnamefont {Corsaro}},
  \ and\ \bibinfo {author} {\bibfnamefont {H.~E.}\ \bibnamefont {Stanley}},\
  }\href@noop {} {\bibfield  {journal} {\bibinfo  {journal} {Proc. Nat. Acc.
  Sci.}\ }\textbf {\bibinfo {volume} {110}},\ \bibinfo {pages} {4899} (\bibinfo
  {year} {2013})}\BibitemShut {NoStop}%
\bibitem [{\citenamefont {Kumar}\ and\ \citenamefont
  {Stanley}(2011)}]{Kumar2011}%
  \BibitemOpen
  \bibfield  {author} {\bibinfo {author} {\bibfnamefont {P.}~\bibnamefont
  {Kumar}}\ and\ \bibinfo {author} {\bibfnamefont {H.~E.}\ \bibnamefont
  {Stanley}},\ }\href@noop {} {\bibfield  {journal} {\bibinfo  {journal} {J.
  Phys. Chem. B}\ }\textbf {\bibinfo {volume} {115}},\ \bibinfo {pages} {14269}
  (\bibinfo {year} {2011})}\BibitemShut {NoStop}%
\bibitem [{\citenamefont {Sciortino}\ \emph {et~al.}(2011)\citenamefont
  {Sciortino}, \citenamefont {Saika-Voivod},\ and\ \citenamefont
  {Poole}}]{Sciortino2011}%
  \BibitemOpen
  \bibfield  {author} {\bibinfo {author} {\bibfnamefont {F.}~\bibnamefont
  {Sciortino}}, \bibinfo {author} {\bibfnamefont {I.}~\bibnamefont
  {Saika-Voivod}}, \ and\ \bibinfo {author} {\bibfnamefont {P.~H.}\
  \bibnamefont {Poole}},\ }\href@noop {} {\bibfield  {journal} {\bibinfo
  {journal} {Phys. Chem. Chem. Phys.}\ }\textbf {\bibinfo {volume} {13}},\
  \bibinfo {pages} {19759} (\bibinfo {year} {2011})}\BibitemShut {NoStop}%
\bibitem [{\citenamefont {Abascal}\ and\ \citenamefont
  {Vega}(2011)}]{Abascal2011}%
  \BibitemOpen
  \bibfield  {author} {\bibinfo {author} {\bibfnamefont {J.~L.~F.}\
  \bibnamefont {Abascal}}\ and\ \bibinfo {author} {\bibfnamefont
  {C.}~\bibnamefont {Vega}},\ }\href@noop {} {\bibfield  {journal} {\bibinfo
  {journal} {J. Chem. Phys.}\ }\textbf {\bibinfo {volume} {134}},\ \bibinfo
  {pages} {186101} (\bibinfo {year} {2011})}\BibitemShut {NoStop}%
\bibitem [{\citenamefont {Paschek}(2004)}]{Paschek2004}%
  \BibitemOpen
  \bibfield  {author} {\bibinfo {author} {\bibfnamefont {D.}~\bibnamefont
  {Paschek}},\ }\href@noop {} {\bibfield  {journal} {\bibinfo  {journal} {J.
  Chem. Phys.}\ }\textbf {\bibinfo {volume} {120}},\ \bibinfo {pages} {6674}
  (\bibinfo {year} {2004})}\BibitemShut {NoStop}%
\bibitem [{\citenamefont {Pi}\ \emph {et~al.}(2009)\citenamefont {Pi},
  \citenamefont {Aragones}, \citenamefont {Vega}, \citenamefont {Noya},
  \citenamefont {Abascal}, \citenamefont {Gonzalez},\ and\ \citenamefont
  {McBride}}]{Pi2009}%
  \BibitemOpen
  \bibfield  {author} {\bibinfo {author} {\bibfnamefont {H.~L.}\ \bibnamefont
  {Pi}}, \bibinfo {author} {\bibfnamefont {J.~L.}\ \bibnamefont {Aragones}},
  \bibinfo {author} {\bibfnamefont {C.}~\bibnamefont {Vega}}, \bibinfo {author}
  {\bibfnamefont {E.~G.}\ \bibnamefont {Noya}}, \bibinfo {author}
  {\bibfnamefont {J.~L.~F.}\ \bibnamefont {Abascal}}, \bibinfo {author}
  {\bibfnamefont {M.~A.}\ \bibnamefont {Gonzalez}}, \ and\ \bibinfo {author}
  {\bibfnamefont {C.}~\bibnamefont {McBride}},\ }\href@noop {} {\bibfield
  {journal} {\bibinfo  {journal} {Mol. Phys.}\ }\textbf {\bibinfo {volume}
  {107}},\ \bibinfo {pages} {365} (\bibinfo {year} {2009})}\BibitemShut
  {NoStop}%
\bibitem [{\citenamefont {Paschek}(2005)}]{Paschek2005}%
  \BibitemOpen
  \bibfield  {author} {\bibinfo {author} {\bibfnamefont {D.}~\bibnamefont
  {Paschek}},\ }\href@noop {} {\bibfield  {journal} {\bibinfo  {journal} {Phys.
  Rev. Lett.}\ }\textbf {\bibinfo {volume} {94}},\ \bibinfo {pages} {217802}
  (\bibinfo {year} {2005})}\BibitemShut {NoStop}%
\bibitem [{\citenamefont {Corradini}\ \emph {et~al.}(2010)\citenamefont
  {Corradini}, \citenamefont {Rovere},\ and\ \citenamefont
  {Gallo}}]{Corradini2010}%
  \BibitemOpen
  \bibfield  {author} {\bibinfo {author} {\bibfnamefont {D.}~\bibnamefont
  {Corradini}}, \bibinfo {author} {\bibfnamefont {M.}~\bibnamefont {Rovere}}, \
  and\ \bibinfo {author} {\bibfnamefont {P.}~\bibnamefont {Gallo}},\
  }\href@noop {} {\bibfield  {journal} {\bibinfo  {journal} {J. Chem. Phys.}\
  }\textbf {\bibinfo {volume} {132}},\ \bibinfo {pages} {134508} (\bibinfo
  {year} {2010})}\BibitemShut {NoStop}%
\bibitem [{\citenamefont {Sciortino}\ \emph {et~al.}(1997)\citenamefont
  {Sciortino}, \citenamefont {Poole}, \citenamefont {Essmann},\ and\
  \citenamefont {Stanley}}]{Sciortino1997}%
  \BibitemOpen
  \bibfield  {author} {\bibinfo {author} {\bibfnamefont {F.}~\bibnamefont
  {Sciortino}}, \bibinfo {author} {\bibfnamefont {P.~H.}\ \bibnamefont
  {Poole}}, \bibinfo {author} {\bibfnamefont {U.}~\bibnamefont {Essmann}}, \
  and\ \bibinfo {author} {\bibfnamefont {H.~E.}\ \bibnamefont {Stanley}},\
  }\href@noop {} {\bibfield  {journal} {\bibinfo  {journal} {Phys. Rev. E}\
  }\textbf {\bibinfo {volume} {55}},\ \bibinfo {pages} {727} (\bibinfo {year}
  {1997})}\BibitemShut {NoStop}%
\bibitem [{\citenamefont {Moore}\ and\ \citenamefont
  {Molinero}(2011)}]{Moore2011}%
  \BibitemOpen
  \bibfield  {author} {\bibinfo {author} {\bibfnamefont {E.~B.}\ \bibnamefont
  {Moore}}\ and\ \bibinfo {author} {\bibfnamefont {V.}~\bibnamefont
  {Molinero}},\ }\href@noop {} {\bibfield  {journal} {\bibinfo  {journal}
  {Nature}\ }\textbf {\bibinfo {volume} {479}},\ \bibinfo {pages} {506}
  (\bibinfo {year} {2011})}\BibitemShut {NoStop}%
\bibitem [{\citenamefont {Limmer}\ and\ \citenamefont
  {Chandler}(2012)}]{Limmer2012}%
  \BibitemOpen
  \bibfield  {author} {\bibinfo {author} {\bibfnamefont {D.~T.}\ \bibnamefont
  {Limmer}}\ and\ \bibinfo {author} {\bibfnamefont {D.}~\bibnamefont
  {Chandler}},\ }\href@noop {} {\bibfield  {journal} {\bibinfo  {journal} {J.
  Chem. Phys.}\ }\textbf {\bibinfo {volume} {137}},\ \bibinfo {pages} {044509}
  (\bibinfo {year} {2012})}\BibitemShut {NoStop}%
\bibitem [{\citenamefont {Huang}\ \emph {et~al.}(2009)\citenamefont {Huang},
  \citenamefont {Wikfeldt}, \citenamefont {Tokushima}, \citenamefont
  {Nordlund}, \citenamefont {Harada}, \citenamefont {Bergmann}, \citenamefont
  {Niebuhr}, \citenamefont {Weiss}, \citenamefont {Horikawa}, \citenamefont
  {Leetmaa}, \citenamefont {Ljungberg}, \citenamefont {Takahashi},
  \citenamefont {Lenz}, \citenamefont {Ojam\"{a}e}, \citenamefont
  {A.~P.~Lyubartsev}, \citenamefont {Pettersson},\ and\ \citenamefont
  {Nilsson}}]{Nilsson2009}%
  \BibitemOpen
  \bibfield  {author} {\bibinfo {author} {\bibfnamefont {C.}~\bibnamefont
  {Huang}}, \bibinfo {author} {\bibfnamefont {K.~T.}\ \bibnamefont {Wikfeldt}},
  \bibinfo {author} {\bibfnamefont {T.}~\bibnamefont {Tokushima}}, \bibinfo
  {author} {\bibfnamefont {D.}~\bibnamefont {Nordlund}}, \bibinfo {author}
  {\bibfnamefont {Y.}~\bibnamefont {Harada}}, \bibinfo {author} {\bibfnamefont
  {U.}~\bibnamefont {Bergmann}}, \bibinfo {author} {\bibfnamefont
  {M.}~\bibnamefont {Niebuhr}}, \bibinfo {author} {\bibfnamefont {T.~M.}\
  \bibnamefont {Weiss}}, \bibinfo {author} {\bibfnamefont {Y.}~\bibnamefont
  {Horikawa}}, \bibinfo {author} {\bibfnamefont {M.}~\bibnamefont {Leetmaa}},
  \bibinfo {author} {\bibfnamefont {M.~P.}\ \bibnamefont {Ljungberg}}, \bibinfo
  {author} {\bibfnamefont {O.}~\bibnamefont {Takahashi}}, \bibinfo {author}
  {\bibfnamefont {A.}~\bibnamefont {Lenz}}, \bibinfo {author} {\bibfnamefont
  {L.}~\bibnamefont {Ojam\"{a}e}}, \bibinfo {author} {\bibfnamefont {S.~S.}\
  \bibnamefont {A.~P.~Lyubartsev}}, \bibinfo {author} {\bibfnamefont
  {L.~G.~M.}\ \bibnamefont {Pettersson}}, \ and\ \bibinfo {author}
  {\bibfnamefont {A.}~\bibnamefont {Nilsson}},\ }\href@noop {} {\bibfield
  {journal} {\bibinfo  {journal} {Proc. Nat. Acc. Sci.}\ }\textbf {\bibinfo
  {volume} {106}},\ \bibinfo {pages} {15214} (\bibinfo {year}
  {2009})}\BibitemShut {NoStop}%
\bibitem [{\citenamefont {Kaya}\ \emph {et~al.}(2013)\citenamefont {Kaya},
  \citenamefont {Schlesinger}, \citenamefont {Yamamoto}, \citenamefont
  {Newberg}, \citenamefont {H.~Bluhm}, \citenamefont {Kendelewicz},
  \citenamefont {G.~E.~Brown}, \citenamefont {Pettersson},\ and\ \citenamefont
  {Nilsson}}]{Nilsson2013}%
  \BibitemOpen
  \bibfield  {author} {\bibinfo {author} {\bibfnamefont {S.}~\bibnamefont
  {Kaya}}, \bibinfo {author} {\bibfnamefont {D.}~\bibnamefont {Schlesinger}},
  \bibinfo {author} {\bibfnamefont {S.}~\bibnamefont {Yamamoto}}, \bibinfo
  {author} {\bibfnamefont {J.~T.}\ \bibnamefont {Newberg}}, \bibinfo {author}
  {\bibfnamefont {H.~O.}\ \bibnamefont {H.~Bluhm}}, \bibinfo {author}
  {\bibfnamefont {T.}~\bibnamefont {Kendelewicz}}, \bibinfo {author}
  {\bibfnamefont {J.}~\bibnamefont {G.~E.~Brown}}, \bibinfo {author}
  {\bibfnamefont {L.~G.~M.}\ \bibnamefont {Pettersson}}, \ and\ \bibinfo
  {author} {\bibfnamefont {A.}~\bibnamefont {Nilsson}},\ }\href@noop {}
  {\bibfield  {journal} {\bibinfo  {journal} {Sci. Rep.}\ }\textbf {\bibinfo
  {volume} {3}},\ \bibinfo {pages} {1074} (\bibinfo {year} {2013})}\BibitemShut
  {NoStop}%
\bibitem [{\citenamefont {Pamuk}\ \emph {et~al.}(2012)\citenamefont {Pamuk},
  \citenamefont {Soler}, \citenamefont {Ramirez}, \citenamefont {Herrero},
  \citenamefont {Stephens}, \citenamefont {Allen},\ and\ \citenamefont
  {Fernandez-Serra}}]{Pamuk12}%
  \BibitemOpen
  \bibfield  {author} {\bibinfo {author} {\bibfnamefont {B.}~\bibnamefont
  {Pamuk}}, \bibinfo {author} {\bibfnamefont {J.~M.}\ \bibnamefont {Soler}},
  \bibinfo {author} {\bibfnamefont {R.}~\bibnamefont {Ramirez}}, \bibinfo
  {author} {\bibfnamefont {C.~P.}\ \bibnamefont {Herrero}}, \bibinfo {author}
  {\bibfnamefont {P.~W.}\ \bibnamefont {Stephens}}, \bibinfo {author}
  {\bibfnamefont {P.~B.}\ \bibnamefont {Allen}}, \ and\ \bibinfo {author}
  {\bibfnamefont {M.-V.}\ \bibnamefont {Fernandez-Serra}},\ }\href@noop {}
  {\bibfield  {journal} {\bibinfo  {journal} {Phys. Rev. Lett.}\ }\textbf
  {\bibinfo {volume} {108}},\ \bibinfo {pages} {193003} (\bibinfo {year}
  {2012})}\BibitemShut {NoStop}%
\bibitem [{\citenamefont {Corsetti}\ \emph {et~al.}(2013)\citenamefont
  {Corsetti}, \citenamefont {Fern\'andez-Serra}, \citenamefont {Soler},\ and\
  \citenamefont {Artacho}}]{Corsetti2013}%
  \BibitemOpen
  \bibfield  {author} {\bibinfo {author} {\bibfnamefont {F.}~\bibnamefont
  {Corsetti}}, \bibinfo {author} {\bibfnamefont {M.-V.}\ \bibnamefont
  {Fern\'andez-Serra}}, \bibinfo {author} {\bibfnamefont {J.~M.}\ \bibnamefont
  {Soler}}, \ and\ \bibinfo {author} {\bibfnamefont {E.}~\bibnamefont
  {Artacho}},\ }\href@noop {} {\bibfield  {journal} {\bibinfo  {journal} {J.
  Phys.: Condens. Matter}\ }\textbf {\bibinfo {volume} {25}},\ \bibinfo {pages}
  {435504} (\bibinfo {year} {2013})}\BibitemShut {NoStop}%
\bibitem [{\citenamefont {Kell}(1977)}]{Kell77}%
  \BibitemOpen
  \bibfield  {author} {\bibinfo {author} {\bibfnamefont {G.~S.}\ \bibnamefont
  {Kell}},\ }\href@noop {} {\bibfield  {journal} {\bibinfo  {journal} {J. Phys.
  Chem. Ref. Data}\ }\textbf {\bibinfo {volume} {6}},\ \bibinfo {pages} {1109}
  (\bibinfo {year} {1977})}\BibitemShut {NoStop}%
\bibitem [{\citenamefont {Ceriotti}\ \emph {et~al.}(2016)\citenamefont
  {Ceriotti}, \citenamefont {Fang}, \citenamefont {Kusalik}, \citenamefont
  {McKenzie}, \citenamefont {Michaelides}, \citenamefont {Morales},\ and\
  \citenamefont {Markland}}]{Markland2016}%
  \BibitemOpen
  \bibfield  {author} {\bibinfo {author} {\bibfnamefont {M.}~\bibnamefont
  {Ceriotti}}, \bibinfo {author} {\bibfnamefont {W.}~\bibnamefont {Fang}},
  \bibinfo {author} {\bibfnamefont {P.~G.}\ \bibnamefont {Kusalik}}, \bibinfo
  {author} {\bibfnamefont {R.~H.}\ \bibnamefont {McKenzie}}, \bibinfo {author}
  {\bibfnamefont {A.}~\bibnamefont {Michaelides}}, \bibinfo {author}
  {\bibfnamefont {M.~A.}\ \bibnamefont {Morales}}, \ and\ \bibinfo {author}
  {\bibfnamefont {T.~E.}\ \bibnamefont {Markland}},\ }\href@noop {} {\bibfield
  {journal} {\bibinfo  {journal} {Chemical Reviews}\ }\textbf {\bibinfo
  {volume} {116}},\ \bibinfo {pages} {7529} (\bibinfo {year}
  {2016})}\BibitemShut {NoStop}%
\bibitem [{\citenamefont {Franks}(2000)}]{Franks}%
  \BibitemOpen
  \bibfield  {author} {\bibinfo {author} {\bibfnamefont {F.}~\bibnamefont
  {Franks}},\ }\href@noop {} {\emph {\bibinfo {title} {Water - A Matrix of
  Life}}}\ (\bibinfo  {publisher} {Royal Society of Chemistry},\ \bibinfo
  {year} {2000})\BibitemShut {NoStop}%
\bibitem [{\citenamefont {R\"{o}ttger}\ \emph {et~al.}(1994)\citenamefont
  {R\"{o}ttger}, \citenamefont {Endriss}, \citenamefont {Ihringer},
  \citenamefont {Doyle},\ and\ \citenamefont {Kuhs}}]{Kuhs94}%
  \BibitemOpen
  \bibfield  {author} {\bibinfo {author} {\bibfnamefont {K.}~\bibnamefont
  {R\"{o}ttger}}, \bibinfo {author} {\bibfnamefont {A.}~\bibnamefont
  {Endriss}}, \bibinfo {author} {\bibfnamefont {J.}~\bibnamefont {Ihringer}},
  \bibinfo {author} {\bibfnamefont {S.}~\bibnamefont {Doyle}}, \ and\ \bibinfo
  {author} {\bibfnamefont {W.~F.}\ \bibnamefont {Kuhs}},\ }\href@noop {}
  {\bibfield  {journal} {\bibinfo  {journal} {Acta Cryst.}\ }\textbf {\bibinfo
  {volume} {B50}},\ \bibinfo {pages} {644} (\bibinfo {year}
  {1994})}\BibitemShut {NoStop}%
\bibitem [{\citenamefont {Allen}(1994)}]{Allen94}%
  \BibitemOpen
  \bibfield  {author} {\bibinfo {author} {\bibfnamefont {P.~B.}\ \bibnamefont
  {Allen}},\ }\href@noop {} {\bibfield  {journal} {\bibinfo  {journal} {Phil.
  Mag. B}\ }\textbf {\bibinfo {volume} {70}},\ \bibinfo {pages} {527} (\bibinfo
  {year} {1994})}\BibitemShut {NoStop}%
\bibitem [{\citenamefont {Habershon}\ \emph {et~al.}(2009)\citenamefont
  {Habershon}, \citenamefont {Markland},\ and\ \citenamefont
  {Manolopoulos}}]{Manolopoulos09}%
  \BibitemOpen
  \bibfield  {author} {\bibinfo {author} {\bibfnamefont {S.}~\bibnamefont
  {Habershon}}, \bibinfo {author} {\bibfnamefont {T.~E.}\ \bibnamefont
  {Markland}}, \ and\ \bibinfo {author} {\bibfnamefont {D.~E.}\ \bibnamefont
  {Manolopoulos}},\ }\href@noop {} {\bibfield  {journal} {\bibinfo  {journal}
  {J. Chem. Phys.}\ }\textbf {\bibinfo {volume} {131}},\ \bibinfo {pages}
  {024501} (\bibinfo {year} {2009})}\BibitemShut {NoStop}%
\bibitem [{\citenamefont {McKenzie}\ \emph {et~al.}(2014)\citenamefont
  {McKenzie}, \citenamefont {Bekker}, \citenamefont {Athokpam},\ and\
  \citenamefont {Ramesh}}]{Ramesh2014}%
  \BibitemOpen
  \bibfield  {author} {\bibinfo {author} {\bibfnamefont {R.~H.}\ \bibnamefont
  {McKenzie}}, \bibinfo {author} {\bibfnamefont {C.}~\bibnamefont {Bekker}},
  \bibinfo {author} {\bibfnamefont {B.}~\bibnamefont {Athokpam}}, \ and\
  \bibinfo {author} {\bibfnamefont {S.~G.}\ \bibnamefont {Ramesh}},\
  }\href@noop {} {\bibfield  {journal} {\bibinfo  {journal} {J. Chem. Phys.}\
  }\textbf {\bibinfo {volume} {140}},\ \bibinfo {pages} {174508} (\bibinfo
  {year} {2014})}\BibitemShut {NoStop}%
\bibitem [{\citenamefont {Li}\ \emph {et~al.}(2011)\citenamefont {Li},
  \citenamefont {Walker},\ and\ \citenamefont {Michaelides}}]{michaelides11}%
  \BibitemOpen
  \bibfield  {author} {\bibinfo {author} {\bibfnamefont {X.-Z.}\ \bibnamefont
  {Li}}, \bibinfo {author} {\bibfnamefont {B.}~\bibnamefont {Walker}}, \ and\
  \bibinfo {author} {\bibfnamefont {A.}~\bibnamefont {Michaelides}},\
  }\href@noop {} {\bibfield  {journal} {\bibinfo  {journal} {Proc. Nat. Acc.
  Sci.}\ }\textbf {\bibinfo {volume} {108}},\ \bibinfo {pages} {6369} (\bibinfo
  {year} {2011})}\BibitemShut {NoStop}%
\bibitem [{\citenamefont {Zeidler}\ \emph {et~al.}(2011)\citenamefont
  {Zeidler}, \citenamefont {Salmon}, \citenamefont {Fischer}, \citenamefont
  {Neuefeind}, \citenamefont {Simonson}, \citenamefont {Lemmel}, \citenamefont
  {Rauch},\ and\ \citenamefont {Markland}}]{Zeidler11}%
  \BibitemOpen
  \bibfield  {author} {\bibinfo {author} {\bibfnamefont {A.}~\bibnamefont
  {Zeidler}}, \bibinfo {author} {\bibfnamefont {P.~S.}\ \bibnamefont {Salmon}},
  \bibinfo {author} {\bibfnamefont {H.~E.}\ \bibnamefont {Fischer}}, \bibinfo
  {author} {\bibfnamefont {J.~C.}\ \bibnamefont {Neuefeind}}, \bibinfo {author}
  {\bibfnamefont {J.~M.}\ \bibnamefont {Simonson}}, \bibinfo {author}
  {\bibfnamefont {H.}~\bibnamefont {Lemmel}}, \bibinfo {author} {\bibfnamefont
  {H.}~\bibnamefont {Rauch}}, \ and\ \bibinfo {author} {\bibfnamefont {T.~E.}\
  \bibnamefont {Markland}},\ }\href@noop {} {\bibfield  {journal} {\bibinfo
  {journal} {Phys. Rev. Lett.}\ }\textbf {\bibinfo {volume} {107}},\ \bibinfo
  {pages} {145501} (\bibinfo {year} {2011})}\BibitemShut {NoStop}%
\bibitem [{\citenamefont {Markland}\ and\ \citenamefont
  {Berne}(2012)}]{Markland2012}%
  \BibitemOpen
  \bibfield  {author} {\bibinfo {author} {\bibfnamefont {T.~E.}\ \bibnamefont
  {Markland}}\ and\ \bibinfo {author} {\bibfnamefont {B.~J.}\ \bibnamefont
  {Berne}},\ }\href@noop {} {\bibfield  {journal} {\bibinfo  {journal}
  {Proceedings of the National Academy of Sciences}\ }\textbf {\bibinfo
  {volume} {109}},\ \bibinfo {pages} {7988} (\bibinfo {year}
  {2012})}\BibitemShut {NoStop}%
\bibitem [{\citenamefont {Romanelli}\ \emph {et~al.}(2013)\citenamefont
  {Romanelli}, \citenamefont {Ceriotti}, \citenamefont {Manolopoulos},
  \citenamefont {Pantalei}, \citenamefont {Senesi},\ and\ \citenamefont
  {Andreani}}]{Ceriotti2013}%
  \BibitemOpen
  \bibfield  {author} {\bibinfo {author} {\bibfnamefont {G.}~\bibnamefont
  {Romanelli}}, \bibinfo {author} {\bibfnamefont {M.}~\bibnamefont {Ceriotti}},
  \bibinfo {author} {\bibfnamefont {D.~E.}\ \bibnamefont {Manolopoulos}},
  \bibinfo {author} {\bibfnamefont {C.}~\bibnamefont {Pantalei}}, \bibinfo
  {author} {\bibfnamefont {R.}~\bibnamefont {Senesi}}, \ and\ \bibinfo {author}
  {\bibfnamefont {C.}~\bibnamefont {Andreani}},\ }\href@noop {} {\bibfield
  {journal} {\bibinfo  {journal} {J. Phys. Chem. Lett.}\ }\textbf {\bibinfo
  {volume} {4}},\ \bibinfo {pages} {3251} (\bibinfo {year} {2013})}\BibitemShut
  {NoStop}%
\bibitem [{\citenamefont {Pamuk}\ \emph {et~al.}(2015)\citenamefont {Pamuk},
  \citenamefont {Allen},\ and\ \citenamefont {Fern\'andez-Serra}}]{Pamuk2015}%
  \BibitemOpen
  \bibfield  {author} {\bibinfo {author} {\bibfnamefont {B.}~\bibnamefont
  {Pamuk}}, \bibinfo {author} {\bibfnamefont {P.~B.}\ \bibnamefont {Allen}}, \
  and\ \bibinfo {author} {\bibfnamefont {M.-V.}\ \bibnamefont
  {Fern\'andez-Serra}},\ }\href@noop {} {\bibfield  {journal} {\bibinfo
  {journal} {Phys. Rev. B}\ }\textbf {\bibinfo {volume} {92}},\ \bibinfo
  {pages} {134105} (\bibinfo {year} {2015})}\BibitemShut {NoStop}%
\bibitem [{\citenamefont {Koichiro~Umemoto}\ and\ \citenamefont
  {Baroni}(2010)}]{Wentzcovitch10}%
  \BibitemOpen
  \bibfield  {author} {\bibinfo {author} {\bibfnamefont {S.~d.~G.}\
  \bibnamefont {Koichiro~Umemoto}, \bibfnamefont {Renata M.~Wentzcovitch}}\
  and\ \bibinfo {author} {\bibfnamefont {S.}~\bibnamefont {Baroni}},\
  }\href@noop {} {\bibfield  {journal} {\bibinfo  {journal} {Chemical Physics
  Letters}\ }\textbf {\bibinfo {volume} {499}},\ \bibinfo {pages} {236}
  (\bibinfo {year} {2010})}\BibitemShut {NoStop}%
\bibitem [{\citenamefont {Murray}\ and\ \citenamefont {Galli}(2012)}]{Galli12}%
  \BibitemOpen
  \bibfield  {author} {\bibinfo {author} {\bibfnamefont {E.~D.}\ \bibnamefont
  {Murray}}\ and\ \bibinfo {author} {\bibfnamefont {G.}~\bibnamefont {Galli}},\
  }\href@noop {} {\bibfield  {journal} {\bibinfo  {journal} {Phys. Rev. Lett.}\
  }\textbf {\bibinfo {volume} {108}},\ \bibinfo {pages} {105502} (\bibinfo
  {year} {2012})}\BibitemShut {NoStop}%
\bibitem [{\citenamefont {Umemoto}\ \emph {et~al.}(2015)\citenamefont
  {Umemoto}, \citenamefont {Sugimura}, \citenamefont {de~Gironcoli},
  \citenamefont {Nakajima}, \citenamefont {Hirose}, \citenamefont {Ohishi},\
  and\ \citenamefont {Wentzcovitch}}]{Wentzcovitch2015}%
  \BibitemOpen
  \bibfield  {author} {\bibinfo {author} {\bibfnamefont {K.}~\bibnamefont
  {Umemoto}}, \bibinfo {author} {\bibfnamefont {E.}~\bibnamefont {Sugimura}},
  \bibinfo {author} {\bibfnamefont {S.}~\bibnamefont {de~Gironcoli}}, \bibinfo
  {author} {\bibfnamefont {Y.}~\bibnamefont {Nakajima}}, \bibinfo {author}
  {\bibfnamefont {K.}~\bibnamefont {Hirose}}, \bibinfo {author} {\bibfnamefont
  {Y.}~\bibnamefont {Ohishi}}, \ and\ \bibinfo {author} {\bibfnamefont {R.~M.}\
  \bibnamefont {Wentzcovitch}},\ }\href@noop {} {\bibfield  {journal} {\bibinfo
   {journal} {Phys. Rev. Lett.}\ }\textbf {\bibinfo {volume} {115}},\ \bibinfo
  {pages} {173005} (\bibinfo {year} {2015})}\BibitemShut {NoStop}%
\bibitem [{\citenamefont {Umemoto}\ and\ \citenamefont
  {Wentzcovitch}(2017)}]{Wentzcovitch2017}%
  \BibitemOpen
  \bibfield  {author} {\bibinfo {author} {\bibfnamefont {K.}~\bibnamefont
  {Umemoto}}\ and\ \bibinfo {author} {\bibfnamefont {R.~M.}\ \bibnamefont
  {Wentzcovitch}},\ }\href {http://stacks.iop.org/1347-4065/56/i=5S3/a=05FA03}
  {\bibfield  {journal} {\bibinfo  {journal} {Japanese Journal of Applied
  Physics}\ }\textbf {\bibinfo {volume} {56}},\ \bibinfo {pages} {05FA03}
  (\bibinfo {year} {2017})}\BibitemShut {NoStop}%
\bibitem [{\citenamefont {Klapproth}\ \emph {et~al.}(2003)\citenamefont
  {Klapproth}, \citenamefont {Goreshnik}, \citenamefont {Staykova},
  \citenamefont {Klein},\ and\ \citenamefont {Kuhs}}]{Kuhs03}%
  \BibitemOpen
  \bibfield  {author} {\bibinfo {author} {\bibfnamefont {A.}~\bibnamefont
  {Klapproth}}, \bibinfo {author} {\bibfnamefont {E.}~\bibnamefont
  {Goreshnik}}, \bibinfo {author} {\bibfnamefont {D.}~\bibnamefont {Staykova}},
  \bibinfo {author} {\bibfnamefont {H.}~\bibnamefont {Klein}}, \ and\ \bibinfo
  {author} {\bibfnamefont {W.~F.}\ \bibnamefont {Kuhs}},\ }\href@noop {}
  {\bibfield  {journal} {\bibinfo  {journal} {Canadian Journal of Physics}\
  }\textbf {\bibinfo {volume} {81}},\ \bibinfo {pages} {503} (\bibinfo {year}
  {2003})}\BibitemShut {NoStop}%
\bibitem [{\citenamefont {Tajima}\ \emph {et~al.}(1982)\citenamefont {Tajima},
  \citenamefont {Matsuo},\ and\ \citenamefont {Suga}}]{Tajima82}%
  \BibitemOpen
  \bibfield  {author} {\bibinfo {author} {\bibfnamefont {Y.}~\bibnamefont
  {Tajima}}, \bibinfo {author} {\bibfnamefont {T.}~\bibnamefont {Matsuo}}, \
  and\ \bibinfo {author} {\bibfnamefont {H.}~\bibnamefont {Suga}},\ }\href@noop
  {} {\bibfield  {journal} {\bibinfo  {journal} {Nature}\ }\textbf {\bibinfo
  {volume} {299}},\ \bibinfo {pages} {810} (\bibinfo {year}
  {1982})}\BibitemShut {NoStop}%
\bibitem [{\citenamefont {Blackman}\ and\ \citenamefont
  {Lisgarten}(1958)}]{Blackman1958}%
  \BibitemOpen
  \bibfield  {author} {\bibinfo {author} {\bibfnamefont {M.}~\bibnamefont
  {Blackman}}\ and\ \bibinfo {author} {\bibfnamefont {N.~D.}\ \bibnamefont
  {Lisgarten}},\ }\href@noop {} {\bibfield  {journal} {\bibinfo  {journal}
  {Advances in Physics}\ }\textbf {\bibinfo {volume} {7}},\ \bibinfo {pages}
  {189} (\bibinfo {year} {1958})}\BibitemShut {NoStop}%
\bibitem [{\citenamefont {Svishchev}\ and\ \citenamefont
  {Kusalik}(1994)}]{Kusalik1994}%
  \BibitemOpen
  \bibfield  {author} {\bibinfo {author} {\bibfnamefont {I.~M.}\ \bibnamefont
  {Svishchev}}\ and\ \bibinfo {author} {\bibfnamefont {P.~G.}\ \bibnamefont
  {Kusalik}},\ }\href@noop {} {\bibfield  {journal} {\bibinfo  {journal} {Phys.
  Rev. Lett.}\ }\textbf {\bibinfo {volume} {239}},\ \bibinfo {pages} {349}
  (\bibinfo {year} {1994})}\BibitemShut {NoStop}%
\bibitem [{\citenamefont {Bertie}\ \emph {et~al.}(1963)\citenamefont {Bertie},
  \citenamefont {D.},\ and\ \citenamefont {Whalley}}]{Bertie1963}%
  \BibitemOpen
  \bibfield  {author} {\bibinfo {author} {\bibfnamefont {J.~E.}\ \bibnamefont
  {Bertie}}, \bibinfo {author} {\bibfnamefont {C.~L.}\ \bibnamefont {D.}}, \
  and\ \bibinfo {author} {\bibfnamefont {E.}~\bibnamefont {Whalley}},\
  }\href@noop {} {\bibfield  {journal} {\bibinfo  {journal} {J. Chem. Phys.}\
  }\textbf {\bibinfo {volume} {38}},\ \bibinfo {pages} {840} (\bibinfo {year}
  {1963})}\BibitemShut {NoStop}%
\bibitem [{\citenamefont {Bertie}\ and\ \citenamefont
  {Whalley}(1964)}]{Bertie1964}%
  \BibitemOpen
  \bibfield  {author} {\bibinfo {author} {\bibfnamefont {J.~E.}\ \bibnamefont
  {Bertie}}\ and\ \bibinfo {author} {\bibfnamefont {E.}~\bibnamefont
  {Whalley}},\ }\href@noop {} {\bibfield  {journal} {\bibinfo  {journal} {J.
  Chem. Phys.}\ }\textbf {\bibinfo {volume} {40}},\ \bibinfo {pages} {1646}
  (\bibinfo {year} {1964})}\BibitemShut {NoStop}%
\bibitem [{\citenamefont {Amaya}\ \emph {et~al.}(2017)\citenamefont {Amaya},
  \citenamefont {Pathak}, \citenamefont {Modak}, \citenamefont {Laksmono},
  \citenamefont {Loh}, \citenamefont {Sellberg}, \citenamefont {Sierra},
  \citenamefont {McQueen}, \citenamefont {Hayes}, \citenamefont {Williams},
  \citenamefont {Messerschmidt}, \citenamefont {Boutet}, \citenamefont {Bogan},
  \citenamefont {Nilsson}, \citenamefont {Stan},\ and\ \citenamefont
  {Wyslouzil}}]{Nilsson2017}%
  \BibitemOpen
  \bibfield  {author} {\bibinfo {author} {\bibfnamefont {A.~J.}\ \bibnamefont
  {Amaya}}, \bibinfo {author} {\bibfnamefont {H.}~\bibnamefont {Pathak}},
  \bibinfo {author} {\bibfnamefont {V.~P.}\ \bibnamefont {Modak}}, \bibinfo
  {author} {\bibfnamefont {H.}~\bibnamefont {Laksmono}}, \bibinfo {author}
  {\bibfnamefont {N.~D.}\ \bibnamefont {Loh}}, \bibinfo {author} {\bibfnamefont
  {J.~A.}\ \bibnamefont {Sellberg}}, \bibinfo {author} {\bibfnamefont {R.~G.}\
  \bibnamefont {Sierra}}, \bibinfo {author} {\bibfnamefont {T.~A.}\
  \bibnamefont {McQueen}}, \bibinfo {author} {\bibfnamefont {M.~J.}\
  \bibnamefont {Hayes}}, \bibinfo {author} {\bibfnamefont {G.~J.}\ \bibnamefont
  {Williams}}, \bibinfo {author} {\bibfnamefont {M.}~\bibnamefont
  {Messerschmidt}}, \bibinfo {author} {\bibfnamefont {S.}~\bibnamefont
  {Boutet}}, \bibinfo {author} {\bibfnamefont {M.~J.}\ \bibnamefont {Bogan}},
  \bibinfo {author} {\bibfnamefont {A.}~\bibnamefont {Nilsson}}, \bibinfo
  {author} {\bibfnamefont {C.~A.}\ \bibnamefont {Stan}}, \ and\ \bibinfo
  {author} {\bibfnamefont {B.~E.}\ \bibnamefont {Wyslouzil}},\ }\href@noop {}
  {\bibfield  {journal} {\bibinfo  {journal} {The Journal of Physical Chemistry
  Letters}\ }\textbf {\bibinfo {volume} {8}},\ \bibinfo {pages} {3216}
  (\bibinfo {year} {2017})}\BibitemShut {NoStop}%
\bibitem [{\citenamefont {Rom\'an-P\'erez}\ \emph {et~al.}(2010)\citenamefont
  {Rom\'an-P\'erez}, \citenamefont {Moaied}, \citenamefont {Soler},\ and\
  \citenamefont {Yndurain}}]{Yndurain10}%
  \BibitemOpen
  \bibfield  {author} {\bibinfo {author} {\bibfnamefont {G.}~\bibnamefont
  {Rom\'an-P\'erez}}, \bibinfo {author} {\bibfnamefont {M.}~\bibnamefont
  {Moaied}}, \bibinfo {author} {\bibfnamefont {J.~M.}\ \bibnamefont {Soler}}, \
  and\ \bibinfo {author} {\bibfnamefont {F.}~\bibnamefont {Yndurain}},\
  }\href@noop {} {\bibfield  {journal} {\bibinfo  {journal} {Phys. Rev. Lett.}\
  }\textbf {\bibinfo {volume} {105}},\ \bibinfo {pages} {145901} (\bibinfo
  {year} {2010})}\BibitemShut {NoStop}%
\bibitem [{\citenamefont {Wang}\ \emph {et~al.}(2011)\citenamefont {Wang},
  \citenamefont {Roman-Perez}, \citenamefont {Soler}, \citenamefont {Artacho},\
  and\ \citenamefont {M.-V.Fernandez-Serra}}]{Jue11}%
  \BibitemOpen
  \bibfield  {author} {\bibinfo {author} {\bibfnamefont {J.}~\bibnamefont
  {Wang}}, \bibinfo {author} {\bibfnamefont {G.}~\bibnamefont {Roman-Perez}},
  \bibinfo {author} {\bibfnamefont {J.~M.}\ \bibnamefont {Soler}}, \bibinfo
  {author} {\bibfnamefont {E.}~\bibnamefont {Artacho}}, \ and\ \bibinfo
  {author} {\bibnamefont {M.-V.Fernandez-Serra}},\ }\href@noop {} {\bibfield
  {journal} {\bibinfo  {journal} {J. Chem. Phys.}\ }\textbf {\bibinfo {volume}
  {134}},\ \bibinfo {pages} {024516} (\bibinfo {year} {2011})}\BibitemShut
  {NoStop}%
\bibitem [{\citenamefont {J.D}\ \emph {et~al.}(1993)\citenamefont {J.D},
  \citenamefont {Kuhs},\ and\ \citenamefont {Finney}}]{Londono1993}%
  \BibitemOpen
  \bibfield  {author} {\bibinfo {author} {\bibfnamefont {L.}~\bibnamefont
  {J.D}}, \bibinfo {author} {\bibfnamefont {W.~F.}\ \bibnamefont {Kuhs}}, \
  and\ \bibinfo {author} {\bibfnamefont {J.~L.}\ \bibnamefont {Finney}},\
  }\href@noop {} {\bibfield  {journal} {\bibinfo  {journal} {J. Chem. Phys.}\
  }\textbf {\bibinfo {volume} {98}},\ \bibinfo {pages} {4878} (\bibinfo {year}
  {1993})}\BibitemShut {NoStop}%
\bibitem [{\citenamefont {Kuhs}\ \emph {et~al.}(1998)\citenamefont {Kuhs},
  \citenamefont {Lobban},\ and\ \citenamefont {Finney}}]{Kuhs1998}%
  \BibitemOpen
  \bibfield  {author} {\bibinfo {author} {\bibfnamefont {W.~F.}\ \bibnamefont
  {Kuhs}}, \bibinfo {author} {\bibfnamefont {C.}~\bibnamefont {Lobban}}, \ and\
  \bibinfo {author} {\bibfnamefont {J.~L.}\ \bibnamefont {Finney}},\
  }\href@noop {} {\bibfield  {journal} {\bibinfo  {journal} {The Review of High
  Pressure Science and Technology}\ }\textbf {\bibinfo {volume} {7}},\ \bibinfo
  {pages} {1141} (\bibinfo {year} {1998})}\BibitemShut {NoStop}%
\bibitem [{\citenamefont {Whalley}\ \emph {et~al.}(1968)\citenamefont
  {Whalley}, \citenamefont {Heath},\ and\ \citenamefont
  {Davidson}}]{Whalley1968}%
  \BibitemOpen
  \bibfield  {author} {\bibinfo {author} {\bibfnamefont {E.}~\bibnamefont
  {Whalley}}, \bibinfo {author} {\bibfnamefont {J.~B.~R.}\ \bibnamefont
  {Heath}}, \ and\ \bibinfo {author} {\bibfnamefont {D.~W.}\ \bibnamefont
  {Davidson}},\ }\href@noop {} {\bibfield  {journal} {\bibinfo  {journal} {J.
  Chem. Phys.}\ }\textbf {\bibinfo {volume} {48}},\ \bibinfo {pages} {2362}
  (\bibinfo {year} {1968})}\BibitemShut {NoStop}%
\bibitem [{\citenamefont {Kamb}(1964)}]{Kamb1964}%
  \BibitemOpen
  \bibfield  {author} {\bibinfo {author} {\bibfnamefont {B.}~\bibnamefont
  {Kamb}},\ }\href@noop {} {\bibfield  {journal} {\bibinfo  {journal} {Acta
  Crystallographica}\ }\textbf {\bibinfo {volume} {17}},\ \bibinfo {pages}
  {1437} (\bibinfo {year} {1964})}\BibitemShut {NoStop}%
\bibitem [{\citenamefont {Kamb}\ \emph {et~al.}(1971)\citenamefont {Kamb},
  \citenamefont {Hamilton}, \citenamefont {LaPlaca},\ and\ \citenamefont
  {Prakash}}]{Kamb1971}%
  \BibitemOpen
  \bibfield  {author} {\bibinfo {author} {\bibfnamefont {B.}~\bibnamefont
  {Kamb}}, \bibinfo {author} {\bibfnamefont {W.~C.}\ \bibnamefont {Hamilton}},
  \bibinfo {author} {\bibfnamefont {S.~J.}\ \bibnamefont {LaPlaca}}, \ and\
  \bibinfo {author} {\bibfnamefont {A.}~\bibnamefont {Prakash}},\ }\href@noop
  {} {\bibfield  {journal} {\bibinfo  {journal} {J. Chem. Phys.}\ }\textbf
  {\bibinfo {volume} {55}},\ \bibinfo {pages} {1934} (\bibinfo {year}
  {1971})}\BibitemShut {NoStop}%
\bibitem [{\citenamefont {Whitworth}\ and\ \citenamefont
  {Petrenko}(1999)}]{icebook}%
  \BibitemOpen
  \bibfield  {author} {\bibinfo {author} {\bibfnamefont {R.~W.}\ \bibnamefont
  {Whitworth}}\ and\ \bibinfo {author} {\bibfnamefont {V.~F.}\ \bibnamefont
  {Petrenko}},\ }\href@noop {} {\emph {\bibinfo {title} {{``Physics of
  ice"}}}}\ (\bibinfo  {publisher} {Oxford University Press},\ \bibinfo {year}
  {1999})\BibitemShut {NoStop}%
\bibitem [{\citenamefont {Salzmann}\ \emph {et~al.}(2009)\citenamefont
  {Salzmann}, \citenamefont {Radaelli}, \citenamefont {Mayer},\ and\
  \citenamefont {Finney}}]{Finney2009}%
  \BibitemOpen
  \bibfield  {author} {\bibinfo {author} {\bibfnamefont {C.~G.}\ \bibnamefont
  {Salzmann}}, \bibinfo {author} {\bibfnamefont {P.~G.}\ \bibnamefont
  {Radaelli}}, \bibinfo {author} {\bibfnamefont {E.}~\bibnamefont {Mayer}}, \
  and\ \bibinfo {author} {\bibfnamefont {J.~L.}\ \bibnamefont {Finney}},\
  }\href@noop {} {\bibfield  {journal} {\bibinfo  {journal} {Phys. Rev. Lett.}\
  }\textbf {\bibinfo {volume} {103}},\ \bibinfo {pages} {105701} (\bibinfo
  {year} {2009})}\BibitemShut {NoStop}%
\bibitem [{\citenamefont {James D.~Jorgensen}\ and\ \citenamefont
  {Worlton}(1984)}]{Worlton84}%
  \BibitemOpen
  \bibfield  {author} {\bibinfo {author} {\bibfnamefont {N.~W.}\ \bibnamefont
  {James D.~Jorgensen}, \bibfnamefont {R.~A.~Beyerlein}}\ and\ \bibinfo
  {author} {\bibfnamefont {T.~G.}\ \bibnamefont {Worlton}},\ }\href@noop {}
  {\bibfield  {journal} {\bibinfo  {journal} {J. Chem. Phys.}\ }\textbf
  {\bibinfo {volume} {81}},\ \bibinfo {pages} {3211} (\bibinfo {year}
  {1984})}\BibitemShut {NoStop}%
\bibitem [{\citenamefont {W.~F.~Kuhs}\ and\ \citenamefont
  {Bliss}(1984)}]{Kuhs84}%
  \BibitemOpen
  \bibfield  {author} {\bibinfo {author} {\bibfnamefont {C.~V.}\ \bibnamefont
  {W.~F.~Kuhs}, \bibfnamefont {J.~L.~Finney}}\ and\ \bibinfo {author}
  {\bibfnamefont {D.~V.}\ \bibnamefont {Bliss}},\ }\href@noop {} {\bibfield
  {journal} {\bibinfo  {journal} {J. Chem. Phys.}\ }\textbf {\bibinfo {volume}
  {81}},\ \bibinfo {pages} {3612} (\bibinfo {year} {1984})}\BibitemShut
  {NoStop}%
\bibitem [{\citenamefont {Nelmes}\ \emph {et~al.}(1993)\citenamefont {Nelmes},
  \citenamefont {Loveday}, \citenamefont {Wilson}, \citenamefont {Besson},
  \citenamefont {Pruzan}, \citenamefont {Klotz},\ and\ \citenamefont
  {\textit{et. al.}}}]{Nelmes1993}%
  \BibitemOpen
  \bibfield  {author} {\bibinfo {author} {\bibfnamefont {R.~J.}\ \bibnamefont
  {Nelmes}}, \bibinfo {author} {\bibfnamefont {J.~S.}\ \bibnamefont {Loveday}},
  \bibinfo {author} {\bibfnamefont {R.~M.}\ \bibnamefont {Wilson}}, \bibinfo
  {author} {\bibfnamefont {J.~M.}\ \bibnamefont {Besson}}, \bibinfo {author}
  {\bibfnamefont {P.}~\bibnamefont {Pruzan}}, \bibinfo {author} {\bibfnamefont
  {S.}~\bibnamefont {Klotz}}, \ and\ \bibinfo {author} {\bibnamefont
  {\textit{et. al.}}},\ }\href@noop {} {\bibfield  {journal} {\bibinfo
  {journal} {Phys. Rev. Lett.}\ }\textbf {\bibinfo {volume} {71}},\ \bibinfo
  {pages} {1192} (\bibinfo {year} {1993})}\BibitemShut {NoStop}%
\bibitem [{\citenamefont {Besson}\ \emph
  {et~al.}(1994{\natexlab{a}})\citenamefont {Besson}, \citenamefont {Pruzan},
  \citenamefont {Klotz}, \citenamefont {Hamel}, \citenamefont {Silvi},
  \citenamefont {Nelmes},\ and\ \citenamefont {\textit{et. al.}}}]{Besson1994}%
  \BibitemOpen
  \bibfield  {author} {\bibinfo {author} {\bibfnamefont {J.~M.}\ \bibnamefont
  {Besson}}, \bibinfo {author} {\bibfnamefont {P.}~\bibnamefont {Pruzan}},
  \bibinfo {author} {\bibfnamefont {S.}~\bibnamefont {Klotz}}, \bibinfo
  {author} {\bibfnamefont {G.}~\bibnamefont {Hamel}}, \bibinfo {author}
  {\bibfnamefont {B.}~\bibnamefont {Silvi}}, \bibinfo {author} {\bibfnamefont
  {R.~J.}\ \bibnamefont {Nelmes}}, \ and\ \bibinfo {author} {\bibnamefont
  {\textit{et. al.}}},\ }\href@noop {} {\bibfield  {journal} {\bibinfo
  {journal} {Phys. Rev. B}\ }\textbf {\bibinfo {volume} {49}},\ \bibinfo
  {pages} {12540} (\bibinfo {year} {1994}{\natexlab{a}})}\BibitemShut {NoStop}%
\bibitem [{\citenamefont {Nelmes}\ \emph {et~al.}(1998)\citenamefont {Nelmes},
  \citenamefont {Loveday}, \citenamefont {Besson}, \citenamefont {Klotz},\ and\
  \citenamefont {Hamel}}]{Nelmes1998}%
  \BibitemOpen
  \bibfield  {author} {\bibinfo {author} {\bibfnamefont {R.~J.}\ \bibnamefont
  {Nelmes}}, \bibinfo {author} {\bibfnamefont {J.~S.}\ \bibnamefont {Loveday}},
  \bibinfo {author} {\bibfnamefont {J.~M.}\ \bibnamefont {Besson}}, \bibinfo
  {author} {\bibfnamefont {S.}~\bibnamefont {Klotz}}, \ and\ \bibinfo {author}
  {\bibfnamefont {G.}~\bibnamefont {Hamel}},\ }\href@noop {} {\bibfield
  {journal} {\bibinfo  {journal} {The Review of High Pressure Science and
  Technology}\ }\textbf {\bibinfo {volume} {7}},\ \bibinfo {pages} {1138}
  (\bibinfo {year} {1998})}\BibitemShut {NoStop}%
\bibitem [{\citenamefont {Umemoto}\ and\ \citenamefont
  {Wentzcovitch}(2004)}]{Wentzcovitch04}%
  \BibitemOpen
  \bibfield  {author} {\bibinfo {author} {\bibfnamefont {K.}~\bibnamefont
  {Umemoto}}\ and\ \bibinfo {author} {\bibfnamefont {R.~M.}\ \bibnamefont
  {Wentzcovitch}},\ }\href@noop {} {\bibfield  {journal} {\bibinfo  {journal}
  {Phys. Rev. B}\ }\textbf {\bibinfo {volume} {69}},\ \bibinfo {pages} {180311}
  (\bibinfo {year} {2004})}\BibitemShut {NoStop}%
\bibitem [{\citenamefont {Umemoto}\ and\ \citenamefont
  {Wentzcovitch}(2005)}]{Wentzcovitch05}%
  \BibitemOpen
  \bibfield  {author} {\bibinfo {author} {\bibfnamefont {K.}~\bibnamefont
  {Umemoto}}\ and\ \bibinfo {author} {\bibfnamefont {R.~M.}\ \bibnamefont
  {Wentzcovitch}},\ }\href@noop {} {\bibfield  {journal} {\bibinfo  {journal}
  {Phys. Rev. B}\ }\textbf {\bibinfo {volume} {71}},\ \bibinfo {pages} {012102}
  (\bibinfo {year} {2005})}\BibitemShut {NoStop}%
\bibitem [{\citenamefont {J-C.~Li}\ and\ \citenamefont
  {Eccleston}(1999)}]{Eccleston99}%
  \BibitemOpen
  \bibfield  {author} {\bibinfo {author} {\bibfnamefont {A.~I.~K.}\
  \bibnamefont {J-C.~Li}, \bibfnamefont {C.~Burnham}}\ and\ \bibinfo {author}
  {\bibfnamefont {R.~S.}\ \bibnamefont {Eccleston}},\ }\href@noop {} {\bibfield
   {journal} {\bibinfo  {journal} {Phys. Rev. B}\ }\textbf {\bibinfo {volume}
  {59}},\ \bibinfo {pages} {9088} (\bibinfo {year} {1999})}\BibitemShut
  {NoStop}%
\bibitem [{\citenamefont {Ziman}(1979)}]{Ziman}%
  \BibitemOpen
  \bibfield  {author} {\bibinfo {author} {\bibfnamefont {J.~M.}\ \bibnamefont
  {Ziman}},\ }\href@noop {} {\emph {\bibinfo {title} {Principles of the Theory
  of Solids}}}\ (\bibinfo  {publisher} {Cambridge University Press},\ \bibinfo
  {year} {1979})\BibitemShut {NoStop}%
\bibitem [{\citenamefont {Ram\'irez}\ \emph {et~al.}(2012)\citenamefont
  {Ram\'irez}, \citenamefont {Neuerburg}, \citenamefont {Fern\'andez-Serra},\
  and\ \citenamefont {Herrero}}]{Herrero12}%
  \BibitemOpen
  \bibfield  {author} {\bibinfo {author} {\bibfnamefont {R.}~\bibnamefont
  {Ram\'irez}}, \bibinfo {author} {\bibfnamefont {N.}~\bibnamefont
  {Neuerburg}}, \bibinfo {author} {\bibfnamefont {M.-V.}\ \bibnamefont
  {Fern\'andez-Serra}}, \ and\ \bibinfo {author} {\bibfnamefont {C.~P.}\
  \bibnamefont {Herrero}},\ }\href@noop {} {\bibfield  {journal} {\bibinfo
  {journal} {J. Chem. Phys.}\ }\textbf {\bibinfo {volume} {137}},\ \bibinfo
  {pages} {044502} (\bibinfo {year} {2012})}\BibitemShut {NoStop}%
\bibitem [{\citenamefont {Salim}\ \emph {et~al.}(2016)\citenamefont {Salim},
  \citenamefont {Willow},\ and\ \citenamefont {Hirata}}]{Hirata2016}%
  \BibitemOpen
  \bibfield  {author} {\bibinfo {author} {\bibfnamefont {M.~A.}\ \bibnamefont
  {Salim}}, \bibinfo {author} {\bibfnamefont {S.~Y.}\ \bibnamefont {Willow}}, \
  and\ \bibinfo {author} {\bibfnamefont {S.}~\bibnamefont {Hirata}},\
  }\href@noop {} {\bibfield  {journal} {\bibinfo  {journal} {The Journal of
  Chemical Physics}\ }\textbf {\bibinfo {volume} {144}},\ \bibinfo {pages}
  {204503} (\bibinfo {year} {2016})}\BibitemShut {NoStop}%
\bibitem [{\citenamefont {Allen}(2015)}]{PhilBulk}%
  \BibitemOpen
  \bibfield  {author} {\bibinfo {author} {\bibfnamefont {P.~B.}\ \bibnamefont
  {Allen}},\ }\href@noop {} {\bibfield  {journal} {\bibinfo  {journal} {Phys.
  Rev. B}\ }\textbf {\bibinfo {volume} {92}},\ \bibinfo {pages} {064106}
  (\bibinfo {year} {2015})}\BibitemShut {NoStop}%
\bibitem [{\citenamefont {Ordej\'on}\ \emph {et~al.}(1996)\citenamefont
  {Ordej\'on}, \citenamefont {Artacho},\ and\ \citenamefont
  {Soler}}]{SiestaPRBRC}%
  \BibitemOpen
  \bibfield  {author} {\bibinfo {author} {\bibfnamefont {P.}~\bibnamefont
  {Ordej\'on}}, \bibinfo {author} {\bibfnamefont {E.}~\bibnamefont {Artacho}},
  \ and\ \bibinfo {author} {\bibfnamefont {J.~M.}\ \bibnamefont {Soler}},\
  }\href@noop {} {\bibfield  {journal} {\bibinfo  {journal} {Phys. Rev. B}\
  }\textbf {\bibinfo {volume} {53}},\ \bibinfo {pages} {10441} (\bibinfo {year}
  {1996})}\BibitemShut {NoStop}%
\bibitem [{\citenamefont {Soler}\ \emph {et~al.}(2002)\citenamefont {Soler},
  \citenamefont {Artacho}, \citenamefont {Gale}, \citenamefont {Garc\'{\i}a},
  \citenamefont {Junquera}, \citenamefont {Ordej\'on},\ and\ \citenamefont
  {S\'anchez-Portal}}]{SiestaJPCM}%
  \BibitemOpen
  \bibfield  {author} {\bibinfo {author} {\bibfnamefont {J.~M.}\ \bibnamefont
  {Soler}}, \bibinfo {author} {\bibfnamefont {E.}~\bibnamefont {Artacho}},
  \bibinfo {author} {\bibfnamefont {J.~D.}\ \bibnamefont {Gale}}, \bibinfo
  {author} {\bibfnamefont {A.}~\bibnamefont {Garc\'{\i}a}}, \bibinfo {author}
  {\bibfnamefont {J.}~\bibnamefont {Junquera}}, \bibinfo {author}
  {\bibfnamefont {P.}~\bibnamefont {Ordej\'on}}, \ and\ \bibinfo {author}
  {\bibfnamefont {D.}~\bibnamefont {S\'anchez-Portal}},\ }\href@noop {}
  {\bibfield  {journal} {\bibinfo  {journal} {J Phys. Condens. Matter}\
  }\textbf {\bibinfo {volume} {14}},\ \bibinfo {pages} {2745} (\bibinfo {year}
  {2002})}\BibitemShut {NoStop}%
\bibitem [{\citenamefont {Perdew}\ \emph {et~al.}(1996)\citenamefont {Perdew},
  \citenamefont {Burke},\ and\ \citenamefont {Ernzerhof}}]{PBE}%
  \BibitemOpen
  \bibfield  {author} {\bibinfo {author} {\bibfnamefont {J.~P.}\ \bibnamefont
  {Perdew}}, \bibinfo {author} {\bibfnamefont {K.}~\bibnamefont {Burke}}, \
  and\ \bibinfo {author} {\bibfnamefont {M.}~\bibnamefont {Ernzerhof}},\
  }\href@noop {} {\bibfield  {journal} {\bibinfo  {journal} {Phys. Rev. Lett.}\
  }\textbf {\bibinfo {volume} {77}},\ \bibinfo {pages} {3865} (\bibinfo {year}
  {1996})}\BibitemShut {NoStop}%
\bibitem [{\citenamefont {Dion}\ \emph {et~al.}(2004)\citenamefont {Dion},
  \citenamefont {Rydberg}, \citenamefont {Schr\"{o}der}, \citenamefont
  {Langreth},\ and\ \citenamefont {Lundqvist}}]{DRSLL}%
  \BibitemOpen
  \bibfield  {author} {\bibinfo {author} {\bibfnamefont {M.}~\bibnamefont
  {Dion}}, \bibinfo {author} {\bibfnamefont {H.}~\bibnamefont {Rydberg}},
  \bibinfo {author} {\bibfnamefont {E.}~\bibnamefont {Schr\"{o}der}}, \bibinfo
  {author} {\bibfnamefont {D.~C.}\ \bibnamefont {Langreth}}, \ and\ \bibinfo
  {author} {\bibfnamefont {B.~I.}\ \bibnamefont {Lundqvist}},\ }\href@noop {}
  {\bibfield  {journal} {\bibinfo  {journal} {Phys. Rev. Lett.}\ }\textbf
  {\bibinfo {volume} {92}},\ \bibinfo {pages} {246401} (\bibinfo {year}
  {2004})}\BibitemShut {NoStop}%
\bibitem [{\citenamefont {Rom\'an-P\'erez}\ and\ \citenamefont
  {Soler}(2009)}]{SolervdW}%
  \BibitemOpen
  \bibfield  {author} {\bibinfo {author} {\bibfnamefont {G.}~\bibnamefont
  {Rom\'an-P\'erez}}\ and\ \bibinfo {author} {\bibfnamefont {J.~M.}\
  \bibnamefont {Soler}},\ }\href@noop {} {\bibfield  {journal} {\bibinfo
  {journal} {Phys. Rev. Lett.}\ }\textbf {\bibinfo {volume} {103}},\ \bibinfo
  {pages} {096102} (\bibinfo {year} {2009})}\BibitemShut {NoStop}%
\bibitem [{\citenamefont {Fanourgakis}\ and\ \citenamefont
  {Xantheas}(2008)}]{ttm3f}%
  \BibitemOpen
  \bibfield  {author} {\bibinfo {author} {\bibfnamefont {G.~S.}\ \bibnamefont
  {Fanourgakis}}\ and\ \bibinfo {author} {\bibfnamefont {S.~S.}\ \bibnamefont
  {Xantheas}},\ }\href@noop {} {\bibfield  {journal} {\bibinfo  {journal} {J.
  Chem. Phys.}\ }\textbf {\bibinfo {volume} {128}},\ \bibinfo {pages} {074506}
  (\bibinfo {year} {2008})}\BibitemShut {NoStop}%
\bibitem [{\citenamefont {Santra}\ \emph {et~al.}(2013)\citenamefont {Santra},
  \citenamefont {Klime\v{s}}, \citenamefont {Tkatchenko}, \citenamefont
  {Alf\`{e}}, \citenamefont {Slater}, \citenamefont {Michaelides},
  \citenamefont {Car},\ and\ \citenamefont {Scheffler}}]{Santra2013}%
  \BibitemOpen
  \bibfield  {author} {\bibinfo {author} {\bibfnamefont {B.}~\bibnamefont
  {Santra}}, \bibinfo {author} {\bibfnamefont {J.}~\bibnamefont {Klime\v{s}}},
  \bibinfo {author} {\bibfnamefont {A.}~\bibnamefont {Tkatchenko}}, \bibinfo
  {author} {\bibfnamefont {D.}~\bibnamefont {Alf\`{e}}}, \bibinfo {author}
  {\bibfnamefont {B.}~\bibnamefont {Slater}}, \bibinfo {author} {\bibfnamefont
  {A.}~\bibnamefont {Michaelides}}, \bibinfo {author} {\bibfnamefont
  {R.}~\bibnamefont {Car}}, \ and\ \bibinfo {author} {\bibfnamefont
  {M.}~\bibnamefont {Scheffler}},\ }\href@noop {} {\bibfield  {journal}
  {\bibinfo  {journal} {The Journal of Chemical Physics}\ }\textbf {\bibinfo
  {volume} {139}},\ \bibinfo {pages} {154702} (\bibinfo {year}
  {2013})}\BibitemShut {NoStop}%
\bibitem [{\citenamefont {Pamuk}(2014)}]{Pamuk2014}%
  \BibitemOpen
  \bibfield  {author} {\bibinfo {author} {\bibfnamefont {B.}~\bibnamefont
  {Pamuk}},\ }\emph {\bibinfo {title} {Nuclear quantum effects in ice phases
  and water from first principles calculations}},\ \href@noop {} {Ph.D.
  thesis},\ \bibinfo  {school} {Stony Brook University} (\bibinfo {year}
  {2014})\BibitemShut {NoStop}%
\bibitem [{\citenamefont {Line}\ and\ \citenamefont
  {Whitworth}(1996)}]{Whitworth1996}%
  \BibitemOpen
  \bibfield  {author} {\bibinfo {author} {\bibfnamefont {C.~M.~B.}\
  \bibnamefont {Line}}\ and\ \bibinfo {author} {\bibfnamefont {R.~W.}\
  \bibnamefont {Whitworth}},\ }\href@noop {} {\bibfield  {journal} {\bibinfo
  {journal} {J. Chem. Phys.}\ }\textbf {\bibinfo {volume} {104}},\ \bibinfo
  {pages} {10008} (\bibinfo {year} {1996})}\BibitemShut {NoStop}%
\bibitem [{\citenamefont {Gagnon}\ \emph {et~al.}(1988)\citenamefont {Gagnon},
  \citenamefont {Kiefte}, \citenamefont {Clouter},\ and\ \citenamefont
  {Whalley}}]{Gagnon88}%
  \BibitemOpen
  \bibfield  {author} {\bibinfo {author} {\bibfnamefont {R.~E.}\ \bibnamefont
  {Gagnon}}, \bibinfo {author} {\bibfnamefont {H.}~\bibnamefont {Kiefte}},
  \bibinfo {author} {\bibfnamefont {M.~J.}\ \bibnamefont {Clouter}}, \ and\
  \bibinfo {author} {\bibfnamefont {E.}~\bibnamefont {Whalley}},\ }\href@noop
  {} {\bibfield  {journal} {\bibinfo  {journal} {J. Chem. Phys.}\ }\textbf
  {\bibinfo {volume} {89}},\ \bibinfo {pages} {4522} (\bibinfo {year}
  {1988})}\BibitemShut {NoStop}%
\bibitem [{\citenamefont {Feistel}\ and\ \citenamefont
  {Wagner}(2006)}]{Wagner06}%
  \BibitemOpen
  \bibfield  {author} {\bibinfo {author} {\bibfnamefont {R.}~\bibnamefont
  {Feistel}}\ and\ \bibinfo {author} {\bibfnamefont {W.}~\bibnamefont
  {Wagner}},\ }\href@noop {} {\bibfield  {journal} {\bibinfo  {journal}
  {Journal of Physical and Chemical Reference Data}\ }\textbf {\bibinfo
  {volume} {35}},\ \bibinfo {pages} {1021} (\bibinfo {year}
  {2006})}\BibitemShut {NoStop}%
\bibitem [{\citenamefont {Hamada}(2010)}]{hamada10}%
  \BibitemOpen
  \bibfield  {author} {\bibinfo {author} {\bibfnamefont {I.}~\bibnamefont
  {Hamada}},\ }\href@noop {} {\bibfield  {journal} {\bibinfo  {journal} {J.
  Chem. Phys.}\ }\textbf {\bibinfo {volume} {133}},\ \bibinfo {pages} {214503}
  (\bibinfo {year} {2010})}\BibitemShut {NoStop}%
\bibitem [{\citenamefont {Gammon}\ \emph {et~al.}(1983)\citenamefont {Gammon},
  \citenamefont {Klefte},\ and\ \citenamefont {Clouter}}]{Gammon83}%
  \BibitemOpen
  \bibfield  {author} {\bibinfo {author} {\bibfnamefont {P.~H.}\ \bibnamefont
  {Gammon}}, \bibinfo {author} {\bibfnamefont {H.}~\bibnamefont {Klefte}}, \
  and\ \bibinfo {author} {\bibfnamefont {M.~J.}\ \bibnamefont {Clouter}},\
  }\href@noop {} {\bibfield  {journal} {\bibinfo  {journal} {The Journal of
  Physical Chemistry}\ }\textbf {\bibinfo {volume} {87}},\ \bibinfo {pages}
  {4025} (\bibinfo {year} {1983})}\BibitemShut {NoStop}%
\bibitem [{\citenamefont {Herrero}\ and\ \citenamefont
  {Ram\'{i}rez}(2011)}]{Ramirez11}%
  \BibitemOpen
  \bibfield  {author} {\bibinfo {author} {\bibfnamefont {C.~P.}\ \bibnamefont
  {Herrero}}\ and\ \bibinfo {author} {\bibfnamefont {R.}~\bibnamefont
  {Ram\'{i}rez}},\ }\href@noop {} {\bibfield  {journal} {\bibinfo  {journal}
  {J. Chem. Phys}\ }\textbf {\bibinfo {volume} {134}},\ \bibinfo {pages}
  {094510} (\bibinfo {year} {2011})}\BibitemShut {NoStop}%
\bibitem [{\citenamefont {Ganeshan}\ \emph {et~al.}(2013)\citenamefont
  {Ganeshan}, \citenamefont {Ram\'irez},\ and\ \citenamefont
  {Fern\'andez-Serra}}]{Sriram2013}%
  \BibitemOpen
  \bibfield  {author} {\bibinfo {author} {\bibfnamefont {S.}~\bibnamefont
  {Ganeshan}}, \bibinfo {author} {\bibfnamefont {R.}~\bibnamefont {Ram\'irez}},
  \ and\ \bibinfo {author} {\bibfnamefont {M.~V.}\ \bibnamefont
  {Fern\'andez-Serra}},\ }\href@noop {} {\bibfield  {journal} {\bibinfo
  {journal} {Phys. Rev. B}\ }\textbf {\bibinfo {volume} {87}},\ \bibinfo
  {pages} {134207} (\bibinfo {year} {2013})}\BibitemShut {NoStop}%
\bibitem [{\citenamefont {Lee}\ \emph {et~al.}(2010)\citenamefont {Lee},
  \citenamefont {Murray}, \citenamefont {Kong}, \citenamefont {Lundqvist},\
  and\ \citenamefont {Langreth}}]{Lee2010}%
  \BibitemOpen
  \bibfield  {author} {\bibinfo {author} {\bibfnamefont {K.}~\bibnamefont
  {Lee}}, \bibinfo {author} {\bibfnamefont {E.~D.}\ \bibnamefont {Murray}},
  \bibinfo {author} {\bibfnamefont {L.}~\bibnamefont {Kong}}, \bibinfo {author}
  {\bibfnamefont {B.~I.}\ \bibnamefont {Lundqvist}}, \ and\ \bibinfo {author}
  {\bibfnamefont {D.~C.}\ \bibnamefont {Langreth}},\ }\href@noop {} {\bibfield
  {journal} {\bibinfo  {journal} {Phys. Rev. B}\ }\textbf {\bibinfo {volume}
  {82}},\ \bibinfo {pages} {081101} (\bibinfo {year} {2010})}\BibitemShut
  {NoStop}%
\bibitem [{\citenamefont {Fortes}\ \emph {et~al.}(2005)\citenamefont {Fortes},
  \citenamefont {Wood}, \citenamefont {Alfredsson}, \citenamefont
  {Vo{\v{c}}adlo},\ and\ \citenamefont {Knight}}]{Fortes2005}%
  \BibitemOpen
  \bibfield  {author} {\bibinfo {author} {\bibfnamefont {A.~D.}\ \bibnamefont
  {Fortes}}, \bibinfo {author} {\bibfnamefont {I.~G.}\ \bibnamefont {Wood}},
  \bibinfo {author} {\bibfnamefont {M.}~\bibnamefont {Alfredsson}}, \bibinfo
  {author} {\bibfnamefont {L.}~\bibnamefont {Vo{\v{c}}adlo}}, \ and\ \bibinfo
  {author} {\bibfnamefont {K.~S.}\ \bibnamefont {Knight}},\ }\href@noop {}
  {\bibfield  {journal} {\bibinfo  {journal} {Journal of Applied
  Crystallography}\ }\textbf {\bibinfo {volume} {38}},\ \bibinfo {pages} {612}
  (\bibinfo {year} {2005})}\BibitemShut {NoStop}%
\bibitem [{\citenamefont {Gagnon}\ \emph {et~al.}(1990)\citenamefont {Gagnon},
  \citenamefont {Kiefte}, \citenamefont {Clouter},\ and\ \citenamefont
  {Whalley}}]{Gagnon1990}%
  \BibitemOpen
  \bibfield  {author} {\bibinfo {author} {\bibfnamefont {R.~E.}\ \bibnamefont
  {Gagnon}}, \bibinfo {author} {\bibfnamefont {H.}~\bibnamefont {Kiefte}},
  \bibinfo {author} {\bibfnamefont {M.~J.}\ \bibnamefont {Clouter}}, \ and\
  \bibinfo {author} {\bibfnamefont {E.}~\bibnamefont {Whalley}},\ }\href@noop
  {} {\bibfield  {journal} {\bibinfo  {journal} {The Journal of Chemical
  Physics}\ }\textbf {\bibinfo {volume} {92}},\ \bibinfo {pages} {1909}
  (\bibinfo {year} {1990})}\BibitemShut {NoStop}%
\bibitem [{\citenamefont {Klotz}\ \emph {et~al.}(2017)\citenamefont {Klotz},
  \citenamefont {Komatsu}, \citenamefont {Kagi}, \citenamefont {Kunc},
  \citenamefont {Sano-Furukawa}, \citenamefont {Machida},\ and\ \citenamefont
  {Hattori}}]{Klotz2017}%
  \BibitemOpen
  \bibfield  {author} {\bibinfo {author} {\bibfnamefont {S.}~\bibnamefont
  {Klotz}}, \bibinfo {author} {\bibfnamefont {K.}~\bibnamefont {Komatsu}},
  \bibinfo {author} {\bibfnamefont {H.}~\bibnamefont {Kagi}}, \bibinfo {author}
  {\bibfnamefont {K.}~\bibnamefont {Kunc}}, \bibinfo {author} {\bibfnamefont
  {A.}~\bibnamefont {Sano-Furukawa}}, \bibinfo {author} {\bibfnamefont
  {S.}~\bibnamefont {Machida}}, \ and\ \bibinfo {author} {\bibfnamefont
  {T.}~\bibnamefont {Hattori}},\ }\href@noop {} {\bibfield  {journal} {\bibinfo
   {journal} {Phys. Rev. B}\ }\textbf {\bibinfo {volume} {95}},\ \bibinfo
  {pages} {174111} (\bibinfo {year} {2017})}\BibitemShut {NoStop}%
\bibitem [{\citenamefont {Ludl}\ \emph {et~al.}(2017)\citenamefont {Ludl},
  \citenamefont {Bove}, \citenamefont {Corradini}, \citenamefont {Saitta},
  \citenamefont {Salanne}, \citenamefont {Bull},\ and\ \citenamefont
  {Klotz}}]{Adriaan2017}%
  \BibitemOpen
  \bibfield  {author} {\bibinfo {author} {\bibfnamefont {A.-A.}\ \bibnamefont
  {Ludl}}, \bibinfo {author} {\bibfnamefont {L.~E.}\ \bibnamefont {Bove}},
  \bibinfo {author} {\bibfnamefont {D.}~\bibnamefont {Corradini}}, \bibinfo
  {author} {\bibfnamefont {A.~M.}\ \bibnamefont {Saitta}}, \bibinfo {author}
  {\bibfnamefont {M.}~\bibnamefont {Salanne}}, \bibinfo {author} {\bibfnamefont
  {C.~L.}\ \bibnamefont {Bull}}, \ and\ \bibinfo {author} {\bibfnamefont
  {S.}~\bibnamefont {Klotz}},\ }\href@noop {} {\bibfield  {journal} {\bibinfo
  {journal} {Phys. Chem. Chem. Phys.}\ }\textbf {\bibinfo {volume} {19}},\
  \bibinfo {pages} {1875} (\bibinfo {year} {2017})}\BibitemShut {NoStop}%
\bibitem [{\citenamefont {Besson}\ \emph
  {et~al.}(1994{\natexlab{b}})\citenamefont {Besson}, \citenamefont {Pruzan},
  \citenamefont {Klotz}, \citenamefont {Hamel}, \citenamefont {Silvi},
  \citenamefont {Nelmes}, \citenamefont {Loveday}, \citenamefont {Wilson},\
  and\ \citenamefont {Hull}}]{Klotz1994}%
  \BibitemOpen
  \bibfield  {author} {\bibinfo {author} {\bibfnamefont {J.~M.}\ \bibnamefont
  {Besson}}, \bibinfo {author} {\bibfnamefont {P.}~\bibnamefont {Pruzan}},
  \bibinfo {author} {\bibfnamefont {S.}~\bibnamefont {Klotz}}, \bibinfo
  {author} {\bibfnamefont {G.}~\bibnamefont {Hamel}}, \bibinfo {author}
  {\bibfnamefont {B.}~\bibnamefont {Silvi}}, \bibinfo {author} {\bibfnamefont
  {R.~J.}\ \bibnamefont {Nelmes}}, \bibinfo {author} {\bibfnamefont {J.~S.}\
  \bibnamefont {Loveday}}, \bibinfo {author} {\bibfnamefont {R.~M.}\
  \bibnamefont {Wilson}}, \ and\ \bibinfo {author} {\bibfnamefont
  {S.}~\bibnamefont {Hull}},\ }\href@noop {} {\bibfield  {journal} {\bibinfo
  {journal} {Phys. Rev. B}\ }\textbf {\bibinfo {volume} {49}},\ \bibinfo
  {pages} {12540} (\bibinfo {year} {1994}{\natexlab{b}})}\BibitemShut {NoStop}%
\bibitem [{\citenamefont {Zhang}\ \emph {et~al.}(2015)\citenamefont {Zhang},
  \citenamefont {Sun}, \citenamefont {Yan}, \citenamefont {Huang},
  \citenamefont {Ma}, \citenamefont {Zou}, \citenamefont {Zheng}, \citenamefont
  {Zhou}, \citenamefont {Gong},\ and\ \citenamefont {Sun}}]{phase}%
  \BibitemOpen
  \bibfield  {author} {\bibinfo {author} {\bibfnamefont {X.}~\bibnamefont
  {Zhang}}, \bibinfo {author} {\bibfnamefont {P.}~\bibnamefont {Sun}}, \bibinfo
  {author} {\bibfnamefont {T.}~\bibnamefont {Yan}}, \bibinfo {author}
  {\bibfnamefont {Y.}~\bibnamefont {Huang}}, \bibinfo {author} {\bibfnamefont
  {Z.}~\bibnamefont {Ma}}, \bibinfo {author} {\bibfnamefont {B.}~\bibnamefont
  {Zou}}, \bibinfo {author} {\bibfnamefont {W.}~\bibnamefont {Zheng}}, \bibinfo
  {author} {\bibfnamefont {J.}~\bibnamefont {Zhou}}, \bibinfo {author}
  {\bibfnamefont {Y.}~\bibnamefont {Gong}}, \ and\ \bibinfo {author}
  {\bibfnamefont {C.~Q.}\ \bibnamefont {Sun}},\ }\href@noop {} {\bibfield
  {journal} {\bibinfo  {journal} {Progress in Solid State Chemistry}\ }\textbf
  {\bibinfo {volume} {43}},\ \bibinfo {pages} {71 } (\bibinfo {year}
  {2015})}\BibitemShut {NoStop}%
\bibitem [{\citenamefont {Kittel}(2005)}]{Kittel}%
  \BibitemOpen
  \bibfield  {author} {\bibinfo {author} {\bibfnamefont {C.}~\bibnamefont
  {Kittel}},\ }\href@noop {} {\emph {\bibinfo {title} {Introduction ot Solid
  State Physics}}}\ (\bibinfo  {publisher} {John Wiley \& Sons, Inc.},\
  \bibinfo {year} {2005})\BibitemShut {NoStop}%
\end{thebibliography}%

\end{document}